\documentclass[12pt]{article}
\usepackage{amsmath,amssymb}
\usepackage{cite}
\numberwithin{equation}{section}

\newcommand{\dis}{\displaystyle}
\newcommand{\ba}{\begin{eqnarray}}
\newcommand{\ea}{\end{eqnarray}}
\newcommand{\non}{\nonumber\\}
\newcommand{\nom}{\nonumber}

\newcommand{\un}[1]{\underline{#1}}

\newcommand{\ket}[1]{\left|{#1}\right\rangle}
\newcommand{\bz}{\mathbb{Z}}
\newcommand{\vth}{\theta}
\newcommand{\vthe}[2]{\theta
\genfrac{[}{]}{0pt}{1}{#1}{#2}}
\newcommand{\soh}{\widehat{\text{so}}}
\newcommand{\lt}{\tilde{\lambda}}
\newcommand{\taub}{\bar{\tau}}
\newcommand{\nn}{\nonumber}
\newcommand{\Zb}{\mathbb{Z}}
\newcommand{\Rb}{\mathbb{R}}

\newcommand{\0}{(0)}
\newcommand{\1}{(1)}
\newcommand{\2}{(2)}
\newcommand{\3}{(3)}
\newcommand{\e}[1]{{\bf e}\!\left[#1\right]}
\DeclareMathOperator*{\Tr}{{\rm Tr}}

\newcommand{\basi}{{\bf bas}}
\newcommand{\vect}{{\bf vec}}
\newcommand{\spi}{{\bf spi}}
\newcommand{\cospi}{{\bf cos}}
\newcommand{\fund}{{\bf fun}}
\newcommand{\cfund}{\overline{\bf fun}}
\newcommand{\so}{{\rm so}}
\newcommand{\SO}{{\rm SO}}
\newcommand{\su}{{\rm su}}
\newcommand{\SU}{\su}
\newcommand{\uone}{{\rm u}(1)}
\newcommand{\isi}{{\bf Ising}}
\newcommand{\tri}{{\bf Tri}}
\newcommand{\pot}{\text{\bf 3-Potts}}
\newcommand{\CFT}{\text{\sf CFT}}
\newcommand{\Ghol}{G_{\rm hol}}

\begin{document}

\thispagestyle{empty}
\begin{flushright}
 \parbox{3.5cm}{KUCP-0193\\{\tt hep-th/0108219}}
\end{flushright}

\vspace*{1cm}
\begin{center}
 {\huge 

 Cascade of Special Holonomy Manifolds\\ and Heterotic String Theory
 }
\end{center}

\vspace*{1cm} 
\begin{center}
 \noindent
 {\large Katsuyuki Sugiyama}

 \vspace{5mm}
 \noindent
 \hspace{0.7cm} \parbox{142mm}{\it
 Department of Fundamental Sciences,
 Faculty of Integrated Human Studies,
 Kyoto University, Yoshida-Nihonmatsu-cho,
 Sakyo-ku, Kyoto 606-8501, Japan.

 E-mail: {\tt sugiyama@phys.h.kyoto-u.ac.jp}
 }

 \vspace{1cm}

 \noindent
 {\large Satoshi Yamaguchi}

 \vspace{5mm}
 \noindent
 \hspace{0.7cm} \parbox{142mm}{\it
 Graduate School of Human and Environmental Studies,
 Kyoto University, Yoshida-Nihonmatsu-cho,
 Sakyo-ku, Kyoto 606-8501, Japan.

 E-mail: {\tt yamaguch@phys.h.kyoto-u.ac.jp}
 }
\end{center}

\vspace{1cm}
\hfill{\bf Abstract\ \ }\hfill\ \\
We investigate heterotic string theory on special holonomy manifolds
including exceptional holonomy $G_2$ and Spin(7) manifolds.  The gauge
symmetry is $F_4$ in a $G_2$ manifold compactification, and $\so(9)$ in
a $Spin(7)$ manifold compactification.  We also study the cascade of the
holonomies: $\so(8)\supset Spin(7)\supset G_2 \supset \su(3) \supset
\su(2)$.  The differences of adjoining groups are described by Ising,
tricritical Ising, 3-state Potts and $\uone$ models. These theories are
essential for spacetime supersymmetries and gauge group enhancements.
As concrete examples, we construct the modular invariant partition
functions and analyze their massless spectra for $G_2$ and $Spin(7)$
orbifolds.  We obtain the relation between topological numbers of the
manifolds and multiplicities of matters in specific representations.

\newpage

\section{Introduction}

It is a long time since the string theory attracted the attention of
particle physicists as a candidate of the unified theory of elementary 
particles and their interactions. 
Extensive works have been devoted to the study of these theories, but 
it seems to be yet out of reach to gain fundamental understanding of
them. 

One of the most important things is the investigation of the
properties of the manifolds on which the string should be
compactified. Particularly the compactifications with minimum 
spacetime supersymmetries have received much attentions. From 
the point of view of the particle physics, 
geometrical properties of internal manifolds determine zero mass
fields in the low energy effective theory and these manifolds play
crucial roles in deciding the phenomenological features of the 
string theories. 

If we require only one spacetime supersymmetry, we need
only one covariantly constant spinor for a fixed chirality and this
leads to the manifolds with minimal numbers of covariantly constant
spinors.
The condition of having an $N=1$ spacetime supersymmetry 
for heterotic string leads to $4$ distinct possibilities for
compactifications namely compactifications 
down to $6$,$4$,$3$,and $2$ dimensions. 
Compactifications to $6$ and $4$ dimensions have been studied 
extensively before 
corresponding to $K3$ and a Calabi-Yau $3$-fold respectively. 
The other two are special cases and 
correspond
to compactification down to $3$ on a $7$ dimensional manifold of
$G_2$ holonomy and compactification down to $2$ on an $8$ dimensional
manifold with $Spin(7)$ holonomy. 
The possible existence of these two special cases had been known for a
long time\cite{BS89,Joyce1,Joyce2,Joyce3}. 
They have been investigated in the 
papers\cite{Shatashvili:1994zw,Acharya:2000gb,Atiyah:2000zz,
Acharya:2001dz,Gibbons:1990ya,Cvetic:2001pg,Edel,Konishi:2001bd,
Cvetic:2001ya,Brandhuber:2001yi,Eguchi:2001xa,Atiyah:2001qf} 
and structures of 
their extended chiral algebras have been clarified\cite{Shatashvili:1994zw}.
The role the $U(1)$ current plays in the 
$N=2$ superconformal theories, is played by tri-critical Ising model
in the case of $G_2$ and Ising model in the case of $Spin(7)$
manifolds.  It is mysterious that
these statistical models appear
unexpectedly in the cases of the exceptional holonomy manifolds. 
One might ask if these phenomena are 
restricted to these scattered exceptional manifolds. 
Also they yield a question that
these manifolds could be related to the 
Calabi-Yau $3$-folds or other 
special holonomy manifolds with 
different dimensions.

The aim of this article is to clarify 
these two questions based on analyses 
of compactifications of string theories. 
The study of spacetime $N=1$ heterotic strings 
is the subject of the present paper.
In the context of string theories the geometries of target manifolds 
can be studied by the worldsheet 
techniques and
conformal field theories
(CFTs) are powerful tools 
in the detailed description of the dynamics.

Motivated with these questions we intend to 
examine toroidal partition functions of the 
heterotic strings by means of CFT techniques.
Particularly we elaborate 
branching rules of
gauge symmetries 
concentrating on the gauge sector of the partition function. 
We find that three $2$ dimensional statistical models, 
``Ising'', ``tricritical Ising'', ``$3$-state Potts'' models 
play important roles in 
connecting $3$ special holonomy manifolds, 
``$Spin(7)$'', ``$G_2$'', ``$CY_3$''. 
At the same time spacetime gauge symmetries are 
respectively enhanced to $SO(9)$, $F_4$, $E_6$ 
in associated heterotic cases for three manifolds.
Then reductions of holonomies about these special 
manifolds are correlated to the enhancements of the 
spacetime gauge symmetries for the $N=1$ susy theories. 
By studying branching rules of characters 
we make clear that 
extra degrees of freedom thrown away under the holonomy reductions 
are transferred to 
those of gauge symmetries and absorbed into 
them as necessary degrees of freedom in their 
enhancements.
At the level of characters of affine Lie algebras 
it might be possible that the $7$ dimensional $G_2$ manifold is 
related with a complex $3$ dimensional Calabi-Yau manifold
by transferring degrees of freedom of 
$3$-state Potts model. 
Then a character of a $U(1)$ current appears and 
we can obtain a state associated with 
a spectral flow operator of the $CY_3$.
It leads us to enhancement of worldsheet currents
from $N=1$ susy to 
$N=2$ susy of the $CY_3$.

The organization of this paper is as follows: In section 2 we will 
review some geometrical facts about manifolds with 
$G_2$ and $Spin(7)$ holonomies and 
associated conformal theories that we will need in the rest of the paper.
We also explain relations between 
ground states of the CFT and 
cohomology classes of the exceptional manifolds.
In section 3 we discuss compactifications of heterotic strings
on exceptional holonomy manifolds. We explain the gauge symmetry
becomes $F_4$ in $G_2$ compactifications, and $\so(9)$ in
$Spin(7)$ compactifications.
In section 4, we study compactifications on special holonomy manifolds
from the point of view of coset CFT of level 1 affine Lie algebras.
We concentrate on characters in the gauge sector of the model and 
study their detailed branching rules. 
We propose cascades 
of special holonomy manifolds with different dimensions 
and they are turned out to be controlled by statistical models.
In section 5 we put 
a review of concrete orbifold examples of these exceptional 
manifolds constructed by Joyce\cite{Joyce1,Joyce2,Joyce3}. 
By combining these left- and right-moving correspondings in the 
heterotic strings, 
we obtain partition functions
of strings compactified 
on $G_2$ and $Spin(7)$ holonomies.
The resulting theories have spacetime $N=1$ supersymmetries. 
For the $G_2$ holonomy case, the tricritical Ising model and $SO(9)$ 
current algebra are combined so that the $3$d spacetime gauge
symmetry is enhanced to an exceptional group $F_4$. 
On the other side the $2$d heterotic string on the 
$Spin(7)$ manifold have an $SO(9)$ spacetime gauge 
symmetry. We would like to point out that the tricritical (or Ising) 
parts are essential for enhancements of 
spacetime gauge symmetries in these $N=1$ theories.
Section 6 is devoted to conclusions and comments. 
In appendix, we collect several useful properties of 
theta functions.

\section{Exceptional Holonomy Manifold}

\subsection{$G_2$ holonomy}

Let us consider a seven manifold $M^{(7)}$ with a $G_2$ 
holonomy. The $G_2$ structure on $M^{(7)}$ is given by a closed 
$G_2$ invariant $3$-form $\Phi$. By including this 
operator, an extended algebra of sigma model on $M^{(7)}$ has been
constructed in the paper\cite{Shatashvili:1994zw} 
based on analyses in the large volume
limit. In addition to a set of stress tensor
$T$ and its superpartner $G$, the conformal algebra contains 
sets of currents $(K,\Phi)$ with spins $(2,3/2)$ 
and $(X,M)$ with spins $(2,5/2)$. 
The $X$ is related with a dual $4$-form $\ast\Phi$ 
and $(X,\Phi)$ is a set of currents $(T^{\tri},G^{\tri})$ of 
$N=1$ additional superconformal algebra.
It is the conformal algebra of the tricritical Ising model with a 
Virasoro central charge $7/10$. Also the theory contains 
a spectral flow operator with the dimension $7/16$, in other words, 
the spin field of the statistical model.
The appearance of the $N=1$ minimal unitary model reflects 
a reduction of holonomy of seven manifold $M^{(7)}$ from $\SO(7)$ to 
$G_2$ and we are left with 
the residual symmetry $\SO(7)/G_2$. 
Its central charge is given as
\ba
\frac{7}{2}-\frac{14}{5}=\frac{7}{10}\,,\nom
\ea
and the correspondence has been proposed 
\ba
\SO(7)/G_2\cong (\mbox{Tricritical Ising})\,.\nom
\ea
Also the original stress tensor $T$ can be decomposed into a sum of 
two commutative Virasoro generators $T=T^{\tri}+T^r$.
The statistical model is a unitary minimal model with central charge 
$c=\frac{7}{10}$. There are $6$ different scaling fields and 
the associated dimensions $h$'s
are listed in table \ref{t1}
\begin{table}[htbp]
 \ba
 &&\begin{array}{|c|cccccc|}\hline
 \mbox{field} & 1 & \epsilon & \epsilon' & \epsilon'' & 
 \sigma & \sigma' \\\hline
 h &  0 & 1/10 &  3/5 &  3/2 &  3/80 &  7/16\\\hline
 \end{array}\nom
\ea
\caption{Conformal dimensions $h$'s of scaling fields in the
tricritical Ising model.}
\label{t1}
\end{table}


The tricritical Ising model is also one of the relevant theories 
endowed with supersymmetry. 
The Neveu-Schwartz sector of the theory contains the 
fields $1$, $\epsilon$, $\epsilon'$, $\epsilon''$. 
In terms of superconformal representations, 
$\epsilon''$ is a descendant of the identity $1$ and 
$\epsilon$ and $\epsilon'$ are superpartners of each other. The fusion
algebra of these $4$ fields closes on itself. 
On the other hand the Ramond sector contains the spin fields 
$\sigma$ and $\sigma' $.
We show the fields assignments in both sectors in table \ref{t2}.
\begin{table}[htbp]
 \ba
 \begin{array}{|c|ccc|}\hline
 h & \mbox{field} & (-1)^F & \mbox{sector}\\\hline
 {}[ 0,3/2 ] & [ 1,\epsilon'' ] & [(+),(-)] & NS\\
 {}[1/10,3/5] & [\epsilon ,\epsilon'] & [(-),(+)] & NS\\
 3/80 & \sigma & (\pm ) & R\\
 7/16 & \sigma' & (\pm ) & R\\\hline
 \end{array}\nom
\ea
\caption{Classification of scaling fields in $N=1$ SCFT. 
The tricritical Ising model can be interpreted as an $N=1$ susy model
in the minimal unitary series.}
\label{t2}
\end{table}

Here we put the ${\bz}_2$ assignments for the tricritical Ising model
according to the paper\cite{Shatashvili:1994zw}. 
In this assignment $(-1)^F=(-1)^{F_I}$ and one can use tricritical
gradings for the whole theory. 
The Ramond ground states are coming in pairs and
the $\pm$ sign reflects this degeneracy and we put two different
$(-1)^F$ assignments in $R$-sectors.

Next we will classify the highest weight representations 
of the algebra by using a set of highest weights $(h^{\tri},h^r)$ 
of $(T^{\tri},T^r)$. These two Virasoro generators are commutative and 
the tricritical Ising part leads to 
unitary highest weight representations of the extended chiral algebra.
Ramond vacua have dimension $\frac{7}{16}$ in this model 
and are classified as
\ba
\begin{array}{rcc}
R; & \dis \ket{\frac{7}{16},0} &\dis \ket{\frac{3}{80},\frac{2}{5}}\,.\nom
\end{array}\nom
\ea
The operator corresponding to the ground state 
$ \ket{\frac{7}{16},0}=\ket{\sigma',0}$ 
plays the role of a spectral flow operator. 
By using fusion relations
\ba
&&\sigma'\cdot\sigma' =1+\epsilon''\,,\non
&&\sigma'\cdot\sigma =\epsilon +\epsilon'\,,\nom
\ea
one can show that 
the Ramond ground state $ \ket{\frac{7}{16},0}=\ket{\sigma',0}$ is 
mapped to an NS vacuum $\ket{0,0}$ and 
the $\ket{\frac{3}{80},\frac{2}{5}}$ is transformed into 
a primary state $\ket{\frac{1}{10},\frac{2}{5}}$ with dimension
$\frac{1}{2}$.
This leads to construct the following states in NS sector
\ba
\begin{array}{rcc}
NS; & \dis \ket{0,0} &\dis \ket{\frac{1}{10},\frac{2}{5}} \,.\nom
\end{array}\nom
\ea

Now we can describe the relation of Ramond ground states with the
cohomology of the manifold $M^{(7)}$.
The target manifold $M^{(7)}$ described by sigma model 
is characterized by its Betti numbers $b_{\ell}$ $(\ell =0,1,\cdots ,7)$
with several relations
\ba
&&b_0=b_7=1\,,\,\,b_1=b_6=0\,,\non
&&b_2=b_5\,,\,\,b_3=b_4\,,\,\nom
\ea
and its Euler number turns out to be $0$. From the point 
of view of geometrical 
consideration, it is known that 
the moduli space ${\cal M}_{geom}$ of the $G_2$ manifold
is related to the structure of the $3$-form $\Phi$ 
and its dimension is given as
\ba
\dim {\cal M}_{geom}=b_3\,.\nom
\ea
In the context of sigma model, the geometrical moduli space is extended 
to a string (CFT) moduli space ${\cal M}_{CFT}$ by 
an antisymmetric $2$-form and its dimension is calculated as
\ba
\dim {\cal M}_{CFT}=b_2+b_3\,.\nom
\ea
In order to see the correspondence with the CFT, 
we glue left- and right-sectors of the 
CFT states and discuss the non-chiral states. 
The relevant states in $(R,R)$ sector are 
constructed as
\ba
\begin{array}{ccc}
\mbox{RR state} & \mbox{number} \\
\dis\ket{(\frac{7}{16},0)_L(\frac{7}{16},0)_R;+} &
 b_0=1\\
\dis\ket{(\frac{3}{80},\frac{2}{5})_L(\frac{3}{80},\frac{2}{5})_R;+} & 
b_2+b_4\\
\dis\ket{(\frac{3}{80},\frac{2}{5})_L(\frac{3}{80},\frac{2}{5})_R;-} & b_3+b_5 \\
\dis\ket{(\frac{7}{16},0)_L(\frac{7}{16},0)_R;-} & b_0=1
\end{array}\nom
\ea
where the signs $\pm$ mean the values of $(-1)^F$. 
Let us consider specific counterparts in the NS sector. 
By acting on Ramond ground state with the 
operator associated with the state
$\ket{(\frac{7}{16},0)_L(\frac{7}{16},0)_R;+}$, 
we obtain $(NS,NS)$ states
\ba
\begin{array}{ccc}
\mbox{NSNS state} & \mbox{number} \\
\dis\ket{(0,0)_L(0,0)_R;+} &
 1\\
\dis\ket{(\frac{1}{10},\frac{2}{5})_L(\frac{1}{10},\frac{2}{5})_R;+} & 
b_2+b_4
\end{array}\nom
\ea
As discussed in the paper\cite{Shatashvili:1994zw}, 
exactly marginal deformations are 
given by operators of the form 
\ba
\begin{array}{cc}
\mbox{state} & \mbox{number} \\
G_{-1/2}\bar{G}_{-1/2}\ket{\dis (\frac{1}{10},\frac{2}{5})_L
(\frac{1}{10},\frac{2}{5})_R;+} & b_2+b_4=b_2+b_3
\end{array}\nom
\ea
which preserve the $G_2$ structure.
These describe string moduli space ${\cal M}_{CFT}$.

\subsection{$Spin(7)$ holonomy}

In this subsection, we will review 
several properties of 
$Spin(7)$ manifold $M^{(8)}$ and its associated 
conformal algebra. Let $M^{(8)}$ be an eight manifold
with a $Spin(7)$ holonomy.
The structure is given by a closed self-dual $Spin(7)$ invariant
$4$-form $\Phi$. The extended symmetry algebra of sigma model 
on $M^{(8)}$ has been found in paper\cite{Shatashvili:1994zw}.
In addition to a set of $N=1$ superconformal currents 
$(T,G)$, it contains operators $(\tilde{X},\tilde{M})$ 
with spins $(2,3/2)$. The set is a pair of an extra
$N=1$ Virasoro conformal algebra $(T_I,G_I)$.
The $\tilde{M}$ corresponds to the Cayley $4$-form $\Phi$ and
the $\tilde{X}$ is the energy momentum tensor 
for the $c=1/2$ Majorana-Weyl fermion (Ising model). 
The latter is related to a spectral flow operator 
with the dimension $1/2$
in the Ising model.
The appearance of this statistical model can 
be explained by a reduction of holonomy for $M^{(8)}$ from 
$\SO(8)$ to $Spin(7)$ 
by calculating central charge of 
$\SO(8)/Spin(7)$
\ba
4-\frac{7}{2}=\frac{1}{2}\,.\nom
\ea
Form this consideration, the correspondence 
has been proposed as
\ba
\SO(8)/Spin(7)\cong (\mbox{Ising model})\,.\nom
\ea
By using this Ising stress tensor $T_I$, 
the original stress tensor can be decomposed into 
a sum of two commutative Virasoro generators
$T=T_I+T_r$. 
This statistical model is a unitary minimal model with central charge 
$c=\frac{1}{2}$. 
There are $3$ local scaling operators in this model: the Ising spin
$\sigma$ and the energy density $\epsilon$ and identity operator $1$.
The associated dimensions $h$'s
are listed in table \ref{t3}
\begin{table}[htbp]
 \ba
 &&\begin{array}{|c|ccc|}\hline
 \mbox{field} & 1 & \sigma & \epsilon \cr\hline
 h & 0 & 1/16 & 1/2 \\
 (-1)^F & (+) & (-) & (+)\\\hline
 \end{array}
 \,\nom
\ea
\caption{Conformal dimensions $h$'s of scaling fields in the Ising model.}
\label{t3}
\end{table}


Here we put the ${\bz}_2$ assignments for the Ising model
according to the paper\cite{Shatashvili:1994zw}. 
In this assignment $(-1)^F=(-1)^{F_I}$ and one can use Ising
gradings for the whole theory. 
It is the ${\bz}_2$ symmetry under spin flips $\sigma\rightarrow
-\sigma$.

Next we will classify our state in the extended algebra by 
a set of two numbers $(h^{\isi},h_r)$:
Ising model highest weight $h^{\isi}$ and the highest weight $h_r$ of the 
$T_r$.
In the Ramond sector we have ground states with dimension $\frac{1}{2}$
and they are classified as
\ba
\begin{array}{rccc}
R; & \ket{\frac{1}{2},0} &\ket{0,\frac{1}{2}} &
\ket{\frac{1}{16},\frac{7}{16}} 
\end{array}\,.\nom
\ea
In this case the state $\ket{\frac{1}{2},0}=\ket{\epsilon ,0}$ 
plays the same role as a spectral flow operator. 
It is nothing but an energy operator of the Ising model. 
By using fusion relations
\ba
\epsilon\cdot \epsilon =1\,,\,\,\epsilon\cdot \sigma =\sigma\,,\nom
\ea
one can show that Ramond ground states 
$\ket{\frac{1}{2},0}$, $\ket{0,\frac{1}{2}}$, 
$\ket{\frac{1}{16},\frac{7}{16}}$ are mapped to respectively NS vacua 
$\ket{0,0}$, $\ket{\frac{1}{2},\frac{1}{2}}$, 
$\ket{\frac{1}{16},\frac{7}{16}}$
\ba
\begin{array}{rccc}
NS; & \ket{0,0} &\ket{\frac{1}{2},\frac{1}{2}} &
\ket{\frac{1}{16},\frac{7}{16}} 
\end{array}\,.\nom
\ea
Now we shall describe the relation of Ramond states with the 
cohomology of the manifold $M^{(8)}$. 
The $Spin(7)$ manifold $M^{(8)}$ associated with this CFT 
is characterized by  geometrical data given by 
Betti numbers $b_{\ell}$ $(\ell =0,1,\cdots ,8)$
together with relations
\ba
&&b_0=b_8=1\,,\,\,b_1=b_7=0\,,\non
&&b_2=b_6\,,\,\,b_3=b_5\,,\non
&&b_3+b_4^{+}-b_2-2b_4^{-}-1=24\,,\nom
\ea
where the $b_4^{\pm}$ mean (anti)self-dual parts 
of the $b_4$.
The Euler number of the eight manifold $M^{(8)}$ is 
calculated as
\ba
\chi =2(b_2-b_3+b_4+1)\,.\nom
\ea
 From the point of view of geometrical 
consideration, it is known that the moduli space ${\cal M}_{geom}$
of the $Spin(7)$ manifold is related to the 
structure of the self-dual $4$-form $\Phi$ and its dimension is given
as
\ba
\dim {\cal M}_{geom}=b_4^{-}+1\,.\nom
\ea
In the context of string theory, the ${\cal M}_{geom}$ is extended 
to a CFT moduli space ${\cal M}_{CFT}$ by $B_{\mu\nu}$ and its
dimension is evaluated as
\ba
\dim {\cal M}_{CFT}=b_2+b_4^{-}+1\,.\nom
\ea
In order to see the correspondence with the 
CFT, we put left- and right-sectors together and discuss 
non-chiral states. 
The relevant states in $(R,R)$  sector 
and associated $(NS,NS)$ counterparts 
are given by the following form
\ba
\begin{array}{ccc}
RR & NSNS & \mbox{number}\\
\ket{\dis (\frac{1}{2},0)_L(\frac{1}{2},0)_R} &
\ket{(0,0)_L(0,0)_R} & b_0=1\\
\ket{\dis (0,\frac{1}{2})_L(0,\frac{1}{2})_R} &
\ket{\dis (\frac{1}{2},\frac{1}{2})_L(\frac{1}{2},\frac{1}{2})_R} &
b_6+b_4^{+}\\
\ket{\dis (\frac{1}{16},\frac{7}{16})_L(\frac{1}{16},\frac{7}{16})_R}
&
\ket{\dis (\frac{1}{16},\frac{7}{16})_L(\frac{1}{16},\frac{7}{16})_R}
& 1+b_2+b_4^{-}\\
\ket{\dis (0,\frac{1}{2})_L(\frac{1}{16},\frac{7}{16})_R}
&
\ket{\dis (\frac{1}{2},\frac{1}{2})_L(\frac{1}{16},\frac{7}{16})_R}
& b_3=b_5\\
\ket{\dis (\frac{1}{16},\frac{7}{16})_L(0,\frac{1}{2})_R}
&
\ket{\dis (\frac{1}{16},\frac{7}{16})_L(\frac{1}{2},\frac{1}{2})_R}
& b_3=b_5
\end{array}\,.\label{sp7}
\ea
The $(R,R)$ and $(NS,NS)$ states are exchanged by the 
operator corresponding to the state $\ket{(\frac{7}{16},0)_L
(\frac{7}{16},0)_R}$. 
As discussed in the paper\cite{Shatashvili:1994zw},
exactly marginal deformations are 
given by operators of the form 
\ba
\begin{array}{cc}
\mbox{State} & \mbox{Number}\\
G_{-1/2}\bar{G}_{-1/2}\ket{\dis (\frac{1}{16},\frac{7}{16})_L
(\frac{1}{16},\frac{7}{16})_R} & 1+b_2+b_4^{-}
\end{array}\nom
\ea
which preserve the $Spin(7)$ structure. 
These describe string moduli space ${\cal M}_{CFT}$.

\section{Compactifications of Heterotic String}

We will consider compactifications of 
$E_8\times E_8$ heterotic string theory\cite{GHMR0,GHMR1,GHMR2} 
on the (real) $D$ dimensional special 
holonomy manifolds.
The resulting theory compactified on $M$ has 
$d$ $(d=10-D)$ dimensional spacetime with 
an $N=1$ supersymmetry and
spacetime gauge symmetries are $G_0\times E_8$.
In order to construct consistent string theories, 
we have to impose several conditions on 
the gauge symmetries.
First of all, let us see these from the 
point of view of worldsheet theories.
In the original $10$ dimensional string,
the left-moving part has an $N=1$ spacetime supersymmetry 
with 
central charge $c=15$ and the right-moving counterpart 
is a bosonic theory with $\bar{c}=26$. 
In quantizing this model, we shall use a
light-cone formula with 
transverse spacetime dimension $(d-2)$ and 
the theory has total central charge
$(c,\bar{c})=(12,24)$.
A spacetime Lorentz group in the light-cone gauge
is $\SO(d-2)$ and is realized 
as level $1$ affine Kac-Moody algebra $\hat{so}(d-2)_1$
by $(d-2)$ free fermions on the worldsheet with central charge
$(d-2)$. 
Similarly the spacetime gauge symmetry $G_0\times E_8$ 
is represented by affine Lie algebras by 
worldsheet gauge fermions with 
central charge $c_{G_0}+8$.
The $D$ dimensional internal part can be described by 
an extended $N=1$ CFT associated with the manifold $M$
in the previous section 
and has central charge $(\frac{3}{2}D,\frac{3}{2}D)$.
By collecting all these parts, we can write down 
conditions of balance of central charges 
on both left- and right-parts
\ba
&&\mbox{right}\,;\,24=(d-2)+\frac{3}{2}D+c_{G_0}+8\,,\non
&&\mbox{left}\,;\,12=\frac{3}{2}(d-2)+\frac{3}{2}D\,.\non
&&\rightarrow d+D=10\,,\,\,c_{G_0}+\frac{1}{2}D=8\,.\nom
\ea
In this article, we will take $M$ as 
exceptional holonomy manifolds $M^{(D)}$ with 
(real) dimension $D$. 
Concrete conditions can be written down for exceptional holonomy cases
\begin{enumerate}
\item[] \underline{G${}_2$ case; $M^{(7)}$}\, \,($d=3$ theory)
\ba
&&12=\frac{3}{2}\times \underline{1}+\frac{3}{2}\times \underline{7}\,,\non
&&24=
\underline{1}+\frac{3}{2}\times \un{7}+(\un{8}+\un{\frac{9}{2}})
\,,\non
&&\qquad \un{1}=d-2=(\mbox{spacetime transverse dimension})\,,\non
&&\qquad \un{7}=(\mbox{dimension of $G_2$ manifold})\,,\non
&&\qquad \un{8}=(\mbox{$c$ of level $1$ affine $E_8$ algebra})=(\mbox{rank of
      $E_8$})\,,\non
&&\qquad \un{\frac{9}{2}}=(\mbox{$c$ of level 1 affine $\SO(9)$ algebra})
\,,\nom
\ea
\item[] \underline{Spin$(7)$ case; $M^{(8)}$}\,\, ($d=2$ theory)
\ba
&&12=\frac{3}{2}\times \underline{0}+\frac{3}{2}\times \underline{8}\,,\non
&&24=\underline{0}+\frac{3}{2}\times \un{8}+(\un{8}+\un{4})
\,,\non
&&\qquad \un{0}=d-2=(\mbox{spacetime transverse dimension})\,,\non
&&\qquad \un{8}=(\mbox{dimension of $Spin(7)$ manifold})\,,\non
&&\qquad \un{8}=(\mbox{$c$ of level 1 affine $E_8$ algebra})=(\mbox{rank of
      $E_8$})\,,\non
&&\qquad \un{4}=(\mbox{$c$ of level 1 affine $\SO(8)$ algebra})
=(\mbox{rank of
      $\SO(8)$})\,.\nom
\ea
\end{enumerate}
The CFTs associated with $M^{(D)}$ have extended algebras 
with spectral flow operators and naive gauge symmetries
$G_0\times E_8$ are enhanced to $G\times E_8$ 
by these special operators.
These operators appear according to reductions of holonomies
from $Spin(7)$ ($\SO(8)$) to $G_2$ ($Spin(7)$) for 
respectively $M^{(7)}$, $M^{(8)}$.
In other words, these are related to 
the degrees of freedom of quotient spaces 
$Spin(7)/G_2$, $\SO(8)/Spin(7)$ and 
turn out to be associated with statistical models 
in the previous section
\ba
G_2&;&Spin(7)/G_2\cong \mbox{Tri-critical
      Ising}\,\,(c=\frac{7}{10})\,,\non
Spin(7)&;&\SO(8)/Spin(7)\cong \mbox{Ising}\,\,
(c=\frac{1}{2})\,.\nom
\ea
By taking account of these 
operators, we can propose 
enhancements of gauge symmetries as
\ba
G_2&;&\frac{7}{10}+\frac{9}{2}=\frac{26}{5}
\rightarrow \{\mbox{tri-critical Ising} \}\times \SO(9)\cong F_4\,,\label{gun1}\\
Spin(7)&;&\frac{1}{2}+4=\frac{9}{2}
\rightarrow \{\mbox{Ising}\}
\times \SO(8)\cong \SO(9)\,,\label{gun2}
\ea
where the left-hand sides of arrows represent 
consistency checks of 
central charges of 
enhanced currents.
In fact, there are embeddings of these gauge groups in $E_8$ 
\ba
&&E_8\supset G_2\times F_4\,,\label{embedding1}\\
&&E_8\supset \SO(7)\times \SO(9)\,,\label{embedding2}
\ea
and this also could be an evidence of the enhancements.

Under these embeddings, the representation $248$ of a visible $E_8$ 
is decomposed by representations of their subgroups.
For the $G_2$ case, 
this decomposition is expressed as
\ba
&&E_8\supset G_2\times F_4\non
&&248=(1,52)\oplus (7,26)\oplus {(14,1)}\,.\nom
\ea
The $7$ of $G_2$ is identified with an index of the 
tangent bundle of the 7 dimensional $G_2$ manifold.
Also each representation of $F_4$ is 
decomposed into representations of $\SO(9)\subset F_4$
\ba
&&26=1+9+16\,,\non
&&52=16+36\,.\nom
\ea
In this article, we consider standard embeddings and 
identify the spin connection of $M^{(7)}$ 
directly with one of gauge $F_4$ singlet fields $(14,1)$.

Next we take the $Spin(7)$ holonomy case.
Through the embedding, the representation $248$ of the $E_8$ 
is decomposed by representations of its subgroups
\ba
&&E_8\supset \SO(7)\times \SO(9)\non
&&248=(1,36)\oplus (7,9)\oplus (8,16)\oplus {(21,1)}\,.\nom
\ea
The $8$ is identified with an index of the tangent space of the 
8 dimensional $Spin(7)$ manifold.
2nd rank antisymmetric tensors $28$ on the 8 dimensional manifold
(with a holonomy $\SO(8)$) are decomposed into self-dual ($\wedge^2_{+}$) 
and anti self-dual parts $(\wedge^2_{-})$
\ba
28=7+21\,.\nom
\ea
It corresponds to a decomposition into 
irreducible $Spin(7)$ modules
\ba
&&\mbox{$2$ form}\,\,\wedge^2({\bf R}^8)\cong \so(8)\,,\non
&&\rightarrow \wedge^2({\bf R}^8)=\wedge^2_{+}\oplus 
\wedge^2_{-}\,,\,\,
\dim \wedge^2_{+}=7\,,\,\,\dim \wedge^2_{-}=21\,,\non
&&\wedge^2_{-}\cong Spin(7)\,.\nom
\ea
In this decomposition, the Cayley $4$-form $\Phi$ plays an important role.
When we regard the $\Phi_{ab}{}^{cd}$ $(a,b,c,d=1,2,\cdots ,8)$ 
as a linear map $\hat{\Phi}$ of an
      $\so(8)$, eigenvalues of the operator $\frac{1}{2}\hat{\Phi}$ turn out to be 
$+1$ or $-3$. According to eigenvalues, we can construct projection
operators  $P_1$, $P_{-3}$
\ba
&&P_1=\frac{3}{4}\left(1+\frac{1}{6}\hat{\Phi}\right)\,,\,\,
P_{-3}=\frac{1}{4}\left(1-\frac{1}{2}\hat{\Phi}\right)\,,\nom
\ea
which project onto $\wedge_{-}^2$, $\wedge_{+}^2$ respectively.
Especially
$Spin(7)$ generators $\hat{G}_{ab}$'s are 
represented as
\ba
&&\hat{G}_{ab}=
\frac{3}{4}\left(\Gamma_{ab}+\frac{1}{6}\Phi_{abcd}\Gamma^{cd}\right)
\in\wedge^2_{-}\,,\non
&&\Gamma_{ab}\,;\,\mbox{$SO(8)$ generator}\,.\nom
\ea
The anti self-dual parts are identified with 
adjoint $21$ and one of them is set equal to 
the spin connection of the $Spin(7)$ manifold
\ba
(21,1)\,.\nom
\ea
The remaining self-dual parts appear
as matters of the vector representations $9$ of $\SO(9)$
\ba
(7,9)\,.\nom
\ea
Also each representation of $\SO(9)$ is 
decomposed into representations of $\SO(8)\subset \SO(9)$
\ba
&&9=1+8_{\vect}\,,\non
&&16=8_{\spi}+8_{\cospi}\,,\non
&&36=8_{\vect}+28\,.\nom
\ea
The subscripts $\vect$, $\spi$, $\cospi$ mean
vector, spinor and cospinor 
representations of $\SO(8)$.

\section{Special holonomy and character relations}

\subsection{Gauge symmetry enhancement from the viewpoint of characters}

Next we study embeddings (\ref{embedding1}),(\ref{embedding2})
 more precisely 
from branching relations of affine Lie algebras.
The gauge symmetries of spacetime are realized by 
affine Kac-Moody algebras on the worldsheet and
we will summarize the properties of several current 
algebras. 
As a first case, we take an affine $\hat{so}(2r)_1$ algebra with 
level $1$. It has central charge $c=r$ and
its spectra (conformal dimension of the primary states)
associated with integrable highest weight representations
are evaluated as
       \begin{align*}
          && \basi:h=0,
          \qquad \vect:h=\frac12,\\
          && \spi:h=\frac{2r}{16},
          \qquad \cospi:h=\frac{2r}{16}.
       \end{align*}
Here, $\basi,\vect,\spi,\cospi$ mean basic, vector, spinor, cospinor
representation respectively.
Also the corresponding characters are evaluated 
by using Jacobi's theta functions
      \begin{align*}
	&\chi^{\so(2r)}_{\basi}
              =\frac12\left(\left(\frac{\theta_3}{\eta}\right)^{r}
              +\left(\frac{\theta_4}{\eta}\right)^{r}\right),\\
	&\chi^{\so(2r)}_{\vect}
              =\frac12\left(\left(\frac{\theta_3}{\eta}\right)^{r}
              -\left(\frac{\theta_4}{\eta}\right)^{r}\right),\\
	&\chi^{\so(2r)}_{\spi}=\chi^{\so(2r)}_{\cospi}
              =\frac12\left(\frac{\theta_2}{\eta}\right)^{r}\,.
       \end{align*}
For the affine $\hat{so}(2r+1)_1$ with level $1$, 
there are three integrable highest weight representations
and the associated conformal dimensions are calculated as
       \begin{align*}
       &\basi:h=0,
          \qquad \vect:h=\frac12,\qquad
          \qquad \spi:h=\frac{2r+1}{16}\,.
       \end{align*}
Their characters are constructed by combining theta functions
       \begin{align*}
	&\chi_{\basi}
              =\frac12\left(\left(\frac{\theta_3}{\eta}\right)^{\frac{2r+1}{2}}
              +\left(\frac{\theta_4}{\eta}\right)^{\frac{2r+1}{2}}\right),\\
	&\chi_{\vect}
              =\frac12\left(\left(\frac{\theta_3}{\eta}\right)^{\frac{2r+1}{2}}
              -\left(\frac{\theta_4}{\eta}\right)^{\frac{2r+1}{2}}\right),\\
	&\chi_{\spi}
              =\frac1{\sqrt2}
        \left(\frac{\theta_2}{\eta}\right)^{\frac{2r+1}{2}}\,.
       \end{align*}
Similarly for level $1$ affine $F_4$, $G_2$, $E_8$ cases, 
we will summarize integrable highest weight representations and their 
conformal dimensions in the 
following lists:
\begin{itemize}
 \item \underline{level 1 affine $F_4$\,\,\,}$(\dis c=\frac{26}{5})$
   \begin{align*}
     &\text{representations\,;\,}\qquad \basi:h=0,\qquad \fund:h=\frac{3}{5}\,,
   \end{align*}
 \item \underline{level 1 affine $G_2$\,\,\,} $(\dis c=\frac{14}{5})$
   \begin{align*}
     &\text{representations\,;\,}\qquad\basi:h=0,\qquad \fund:h=\frac{2}{5}\,,
   \end{align*}
 \item \underline{level 1 affine $E_8$\,\,\,} $(c=8)$
   \begin{align*}
     &\text{representations\,;\,}\qquad \basi:h=0\,,
   \end{align*}
\end{itemize}
where the ``$\basi$'', ``$\fund$'' represent respectively 
the basic, fundamental representations of the 
corresponding algebras.
Under these preparations we can obtain the tricritical Ising model by 
the coset construction 
$(\hat{F}_4)_1/\hat{so}(9){}_1$.
Then branching relation is expressed in the characters of each 
CFT algebra
   \begin{align*}
      \chi^{(F_4)}_{\Lambda}=\sum_{\lambda}\chi^{
    \tri}_{\Lambda,\lambda}\; \chi^{\so(9)}_{\lambda}\,.
   \end{align*}
The symbol $\Lambda$ ($=$ \basi, \fund) expresses each highest weight
representation of $(\hat{F}_4)_1$ 
and $\lambda$ ($=$ \basi, \vect, \spi) 
       labels $\hat{so}(9)_1$ counterparts.
The conformal dimensions of the Verma modules
        $(\Lambda,\lambda)$ are evaluated in the following table
\begin{center}
\begin{tabular}{|c||c|c|c|c|c|c|}\hline
$(\Lambda,\lambda)$ &(\basi,\basi) &(\basi,\vect) 
&(\basi,\spi) &(\fund,\basi) &(\fund,\vect) &(\fund,\spi) \\ \hline
$h$ &0 &3/2 &7/16 &3/5 &1/10 &3/80 \\ \hline
\end{tabular}\,.
\end{center}
That is to say, the $\hat{F}_4$ characters are 
decomposed according to the highest weights of 
the tricritical Ising model in the following way
   \begin{align}
    &\chi^{F_4}_{\basi}=\chi^{\tri}_{0}\chi^{\so(9)}_{\basi}
          +\chi^{\tri}_{3/2}\chi^{\so(9)}_{\vect}
          +\chi^{\tri}_{7/16}\chi^{\so(9)}_{\spi}\,,\nn\\
    &\chi^{F_4}_{\fund}=\chi^{\tri}_{3/5}\chi^{\rm \so(9)}_{\basi}
          +\chi^{\tri}_{1/10}\chi^{\so(9)}_{\vect}
          +\chi^{\tri}_{3/80}\chi^{\so(9)}_{\spi}\,.
   \label{F4/so(9)}
   \end{align}
These are nothing but concrete realizations of 
the enhancement of gauge symmetry in Eq.(\ref{gun1}) from $\SO(9)$ to $F_4$.
Also similar decompositions can be performed for \\ sets $\{\hat{so}(7)_1,
(\hat{G}_2)_1 ,(\mbox{tricritical Ising})\}$ and 
$\{(\hat{E}_8)_1 ,\hat{so}(7)_1 ,\hat{so}(9)_1\}$
by applying the same technique as the $\hat{F}_4$ case
   \begin{align}
    &\chi^{\so(7)}_{\basi}=\chi^{\tri}_{0}\chi^{G_2}_{\basi}
                         +\chi^{\tri}_{3/5}\chi^{G_2}_{\fund}\,,\nn\\
    &\chi^{\so(7)}_{\vect}=\chi^{\tri}_{3/2}\chi^{G_2}_{\basi}
                         +\chi^{\tri}_{1/10}\chi^{G_2}_{\fund}\,,\nn\\
    &\chi^{\so(7)}_{\spi}=\chi^{\tri}_{7/16}\chi^{G_2}_{\basi}
                         +\chi^{\tri}_{3/80}\chi^{G_2}_{\fund}\,,
   \label{so(7)/G2}\\
    &\chi^{E_8}_{\basi}=\chi^{\so(7)}_{\basi}\chi^{\so(9)}_{\basi}+
                    \chi^{\so(7)}_{\vect}\chi^{\so(9)}_{\vect}+
                    \chi^{\so(7)}_{\spi}\chi^{\so(9)}_{\spi}\,.\nn
   \end{align}
The first three equations mean that the 
holonomy group $G_2$ of our manifold $M$ is embedded in the $Spin(7)$.
Collecting all these relations we can show an equation  
among characters for $(\hat{E}_8)_1$, $(\hat{F}_4)_1$ and
$(\hat{G}_2)_1$
   \begin{align*}
    \chi^{E_8}_{\basi}
      =\chi^{F_4}_{\basi}\chi^{G_2}_{\basi}
              +\chi^{F_4}_{\fund}\chi^{G_2}_{\fund}\,.
   \end{align*}
This describes embeddings of gauge groups $G_2$ and $F_4$ in $E_8$ 
considered in Eq.(\ref{embedding1}).
The degrees of freedom in the tricritical Ising model are included in
the symmetry algebra $\soh (7)$. But they are transferred from this
$\soh (7)$
to $\soh (9)$ and enhance the spacetime gauge symmetry from 
$\SO(9)$ to $F_4$. 

Next we investigate the $Spin(7)$ holonomy case by taking account of
the coset construction $\soh$$(9)_1/\soh$$(8)_1$ of the Ising model.
The branching relation is expressed by characters of these CFTs
\begin{align*}
 \chi^{\so(9)}_{\Lambda}=
 \sum_{\lambda}\chi^{\isi}_{\Lambda,\lambda}\chi^{\so(8)}_{\lambda}\,.
\end{align*}
Here the $\Lambda$ expresses highest weight representations of
$\hat{so}(9)$ 
and $\lambda$ labels $\hat{so}(8)$ representations. 
The conformal dimensions of the Verma modules $(\Lambda ,\lambda)$ are 
evaluated in the following table
\begin{center}
\begin{tabular}{|c||c|c|c|}\hline
$(\Lambda,\lambda)$ &(\basi,\basi), (\vect,\vect) &(\basi,\vect), (\vect,\basi) &(\spi,\spi), (\spi,\cospi) \\ \hline
$h$ &0 &1/2 &1/16 \\ \hline
\end{tabular}
\end{center}
and we can write down decompositions of characters of $\hat{so}(9)_1$
in terms of the weights of the Ising model concretely
   \begin{align}
    &\chi^{\so(9)}_{\basi}=\chi^{\isi}_{0}\chi^{\so(8)}_{\basi}
                         +\chi^{\isi}_{1/2}\chi^{\so(8)}_{\vect}\,,\nn\\
    &\chi^{\so(9)}_{\vect}=\chi^{\isi}_{1/2}\chi^{\so(8)}_{\basi}
                         +\chi^{\isi}_{0}\chi^{\so(8)}_{\vect}\,,\nn\\
    &\chi^{\so(9)}_{\spi}=\chi^{\isi}_{1/16}\chi^{\so(8)}_{\spi}
                         +\chi^{\isi}_{1/16}\chi^{\so(8)}_{\cospi}\,.
   \label{so(9)/so(8)}
   \end{align}
These show the enhancement of gauge symmetry 
in Eq.(\ref{gun2}) from $\SO(8)$ to $\SO(9)$. 
On the other hand the holonomy $Spin(7)$ is embedded in the $\SO(8)$ 
and this fact leads us to relations among characters of $\soh$$(7)$ and
$\soh$$(8)$
   \begin{align}
    &\chi^{\so(8)}_{\basi}=\chi^{\isi}_{0}\chi^{\so(7)}_{\basi}
                         +\chi^{\isi}_{1/2}\chi^{\so(7)}_{\vect}\,,\nn\\
    &\chi^{\so(8)}_{\vect}=\chi^{\isi}_{1/2}\chi^{\so(7)}_{\basi}
                         +\chi^{\isi}_{0}\chi^{\so(7)}_{\vect}\,,\nn\\
    &\chi^{\so(8)}_{\spi}=\chi^{\isi}_{1/16}\chi^{\so(7)}_{\spi}\,,\,\,
    \chi^{\so(8)}_{\cospi}=\chi^{\isi}_{1/16}\chi^{\so(7)}_{\spi}\,.
    \label{so(8)/so(7)}
   \end{align}
By gathering these equations
together with a decomposition of the $E_8$ character in terms 
of $\so (8)$'s
   \begin{align*}
    \chi^{E_8}_{\basi}&=\chi^{\so(8)}_{\basi}\chi^{\so(8)}_{\basi}+
                    \chi^{\so(8)}_{\vect}\chi^{\so(8)}_{\vect}\\
                   &\qquad  +\chi^{\so(8)}_{\spi}\chi^{\so(8)}_{\spi}+
                    \chi^{\so(8)}_{\cospi}\chi^{\so(8)}_{\cospi}\,,
   \end{align*}
we conclude the branching relation in terms of $\so(7)$ and $\so(9)$
  \begin{align*}
    \chi^{E_8}_{\basi}&=\chi^{\so(7)}_{\basi}\chi^{\so(9)}_{\basi}+
                    \chi^{\so(7)}_{\vect}\chi^{\so(9)}_{\vect}+
                    \chi^{\so(7)}_{\spi}\chi^{\so(9)}_{\spi}\,.
   \end{align*}
This describes embeddings of gauge groups $\SO(7)$ and $\SO(9)$ into
$E_8$ 
in Eq.(\ref{embedding2}). In this case the degrees of freedom in the
Ising model are transferred from one $\soh (8)$ to the other $\soh (8)$ 
and spacetime gauge symmetry is enhanced from $\SO(8)$ to $\SO(9)$. 
At the same time, the holonomy of $M^{(8)}$ is reduced from $\SO(8)$ to 
$Spin(7)$. 
It is amazing that these phenomena about holonomies and 
gauge symmetries can be explained rigorously at the 
level of affine Lie algebras.

\subsection{Relation to Calabi-Yau 3-fold and K3 compactification}
Let us consider 8-dimensional space $M^{(8)}_{0}$ which is the whole
transverse space of the string theory in light-cone gauge. $M^{(8)}_{0}$
might be a compact manifold, or a direct product ${\Rb}^{(D-2)}\times
M^{(10-D)}$, where $M^{(10-D)}$ is a $(10-D)$-dimensional compact manifold.

Generally, the holonomy group $\Ghol$ of $M^{(8)}_{0}$ is included in
$\so(8)$. In this case, the manifest gauge symmetry of the heterotic
string theory on $M^{(8)}_0$ is $\so(8)$, and there may be no 
supersymmetries. For this reason, let us denote this class of whole
8-dimensional (Ricci flat) manifolds or sigma models on these manifolds
as $\CFT(\so(8))$.

As a subset of $\CFT(\so(8))$, we consider manifolds with holonomy
group $\Ghol\subset Spin(7)\subset \so(8)$.  We will name this class of
manifolds or sigma model on them as $\CFT(\so(7))$.  In such theories,
the difference between \so(8) and \so(7) (in coset CFT meaning) ---
Ising model --- is not broken by the holonomy.  The relation
\so(8)$/$\so(7) $=$ (Ising) is shown in Eqs.(\ref{so(8)/so(7)}).  This
extra ``symmetry'' causes the spacetime supersymmetry, and makes the
naive gauge symmetry \so(8) enhanced to \so(9) as in 
Eqs.(\ref{so(9)/so(8)}).  This relation can be expressed by Eqs.
(\ref{so(8)/so(7)}). 

There is a certain subset of $\CFT(\so(7))$ which has more
spacetime supersymmetry (the number of supercharges is larger) and larger
gauge symmetry.  Its holonomy $\Ghol$ is included in
$G_2\subset\so(7)$. We call this class of manifolds $\CFT(G_2)$. The
prime example in this class of manifolds are a direct product of a
flat line and a 7-dimensional $G_2$ holonomy manifold. A theory in
$\CFT(G_2)$ has more supercharges in spacetime and larger gauge
symmetry than a general element in $\CFT(\so(7))$ because the theory in
$\CFT(G_2)$ has an extra symmetry expressed by the coset theory
$\so(7)/G_2\cong$ (tricritical Ising model). The relation
$\so(7)/G_2\cong$ (tricritical Ising model) is shown in
Eqs.(\ref{so(7)/G2}). This tricritical Ising model causes 
more supercharges, and larger gauge group than the general theory in
$\CFT(\so(7))$ has. For example, the gauge symmetry enhancement $\so(9)$ to
$F_4$ occurs when we combine the $\so(9)$ algebra and the 
tricritical Ising model as shown in Eqs.(\ref{F4/so(9)}).

Moreover, as a subset of $\CFT(G_2)$, we can consider a class of 
manifolds whose holonomies are included in $\su(3)\subset G_2$.  We call
this class of manifolds as $\CFT(\su(3))$. The prime example we mainly
consider is a direct product of flat $\Rb^2$ and a Calabi-Yau 3-fold.
A theory in $\CFT(\su(3))$ has more supercharges in spacetime and
larger gauge symmetry than the general theory in $\CFT(G_2)$ because the
theory in $\CFT(\su(3))$ has the extra symmetry expressed by the coset
$G_2/\su(3)$. This coset $G_2/\su(3)$ turns out to be the 3-state
Potts model from the following relations
   \begin{align}
    \chi^{G_2}_{\basi}=C^{\pot}_{0}\chi^{\su(3)}_{\basi}+
              C^{\pot}_{2/3}\chi^{\su(3)}_{\fund}+
              C^{\pot}_{2/3}\chi^{\su(3)}_{\cfund},\nn\\
    \chi^{G_2}_{\fund}=C^{\pot}_{2/5}\chi^{\rm \su(3)}_{\basi}+
              C^{\pot}_{1/15}\chi^{\su(3)}_{\fund}+
              C^{\pot}_{1/15}\chi^{\su(3)}_{\cfund}.
     \label{G2/su(3)}
   \end{align}
By the effect of this 3-state Potts model, a theory in $\CFT(\su(3))$
has the larger gauge symmetry $E_6$ than the gauge symmetry
$F_4$ of a general theory in $\CFT(G_2)$. It can be shown as branching rules
   \begin{align}
    \chi^{E_6}_{\basi}=C^{\pot}_{0}\chi^{F_4}_{\basi}
             + C^{\pot}_{2/5}\chi^{F_4}_{\fund},\nn\\
    \chi^{E_6}_{\fund}=\chi^{E_6}_{\cfund}=C^{\pot}_{2/3}
                \chi^{F_4}_{\basi}
             + C^{\pot}_{1/15}\chi^{F_4}_{\fund}.
   \label{E6/F4}
   \end{align}
A theory in $\CFT(\su(3))$ also has more supercharges in spacetime 
 than a general theory in $\CFT(G_2)$. 
This theory also has a peculiar property.
The $N=1$ theory on the worldsheet has ${\Zb}_2$ 
$R$-symmetry. 
But $R$-symmetry of 
the theory in this $\CFT(\su(3))$
is continuous $U(1)$ and this theory
has worldsheet
 $N=2$ supersymmetry.

As a subset of $\CFT(\su(3))$, we can consider a class
of manifolds (or CFT) $\CFT(\su(2))\subset\CFT(\su(3))$. 
A manifold in this class has a holonomy included in $\su(2)\subset\su(3)$.
The prime example of the manifold in $\CFT(\su(2))$ 
is K3$\times \Rb^4$, which we mainly consider.
The difference between a theory in $\CFT(\su(2))$ and a
general theory in $\CFT(\su(3))$
is evaluated by the coset 
$\su(3)/\su(2)\cong \uone_{3}$ (see also appendix. \ref{appendix-WZW}).
That is seen from relations
   \begin{align}
    \chi^{\su(3)}_{\basi}=\frac{\Theta_{0,3}}{\eta}\chi^{\su(2)}_{\basi}
                 +\frac{\Theta_{3,3}}{\eta}\chi^{\su(2)}_{\fund},\nn\\
    \chi^{\su(3)}_{\fund}=\chi^{\su(3)}_{\cfund}
                =\frac{\Theta_{2,3}}{\eta}\chi^{\su(2)}_{\basi}
                 +\frac{\Theta_{1,3}}{\eta}\chi^{\su(2)}_{\fund}.
   \label{su(3)/su(2)}
   \end{align}
A theory in $\CFT(\su(2))$ has the larger gauge symmetry $E_7$
than the $E_6$ of
a general theory in $\CFT(\su(3))$ by this $\uone_3$.
This is seen from equations about characters
\begin{align}
& \chi^{E_7}_{\basi}=\frac{\Theta_{0,3}}{\eta}\chi^{E_6}_{\basi}
                   +\frac{\Theta_{2,3}}{\eta}\chi^{E_6}_{\fund}
                   +\frac{\Theta_{2,3}}{\eta}\chi^{E_6}_{\cfund},\nn\\
& \chi^{E_7}_{\fund}=\frac{\Theta_{3,3}}{\eta}\chi^{E_6}_{\basi}
                   +\frac{\Theta_{1,3}}{\eta}\chi^{E_6}_{\fund}
                   +\frac{\Theta_{1,3}}{\eta}\chi^{E_6}_{\cfund}.
   \label{E7/E6}
\end{align}
The number of spacetime supercharges of a theory in $\CFT(\su(2))$ is
also larger than that of a general theory in $\CFT(\su(3))$.

Finally, there is a class of flat manifolds, such as $\Rb^8$.
We name this class $\CFT(1)\subset \CFT(\su(2))$. 
A flat CFT in $\CFT(1)$ has more spacetime supercharges
and the larger gauge symmetry than a general CFT in $\CFT(\su(2))$
because a flat CFT has the extra symmetry $\su(2)$. 
A theory in $\CFT(1)$ has the largest spacetime
supersymmetry, and the largest gauge group $E_8$. The gauge symmetry
enhancement from $E_7$ to $E_8$ can be seen from the relation 
about characters
\begin{align*}
 & \chi^{E_8}_{\basi}=\chi^{\su(2)}_{\basi}\chi^{E_7}_{\basi}
                   +\chi^{\su(2)}_{\fund}\chi^{E_7}_{\fund}.
\end{align*}
Collecting these results, we find a sequence of inclusions of
holonomy groups
\begin{align}
  {\so(8)}\supset{\so(7)}\supset {G_2}
       \supset{\su(3)}\supset{\su(2)}\supset \{1\} .
 \label{seq-holonomy}
\end{align}
This induces a sequence of classes of manifolds (or theories)
\begin{align}
 \CFT(\so(8))\supset\CFT(\so(7))\supset\CFT(G_2)\supset
\CFT(\su(3))\supset\CFT(\su(2))\supset\CFT(1).
 \label{seq-theory}
\end{align}
On the other side, there is also a sequence of gauge groups of theories
\begin{align}
 {\so(8)}\subset{\so(9)}\subset{F_4}\subset{E_6}\subset{E_7}\subset{E_8}.
 \label{seq-gauge}
\end{align}
Each gauge group corresponds to each theory associated with a specific
holonomy manifold.
We can describe coset CFTs of subsequent two theories as rational CFTs
\begin{align}
& { \so(8)}/{ \so(7)}\cong { \so(9)}/{ \so(8)}
       \cong {\text{Ising}}  ,\qquad
 { \so(7)}/{G_2}\cong {F_4}/{ \so(9)}
       \cong {\text{tricritical Ising}}  ,\label{coset1}\\
& {G_2}/{\su(3)}\cong {E_6}/{F_4}
       \cong {\text{3-state Potts}}  ,\qquad
 {\su(3)}/{\su(2)}\cong {E_7}/{E_6}
       \cong {\uone_3} , \label{coset2}\\
& {\su(2)}/\{1\}\cong {E_8}/{E_7}
       \cong {\su(2)}  .\nn
\end{align}
These quotient theories play essential roles in gauge group enhancements
and understanding spacetime supersymmetries.

Also, these sequences can be used to analyze special holonomy
manifolds. For example, when one intends to study Calabi-Yau
compactifications, he should consider the decomposition of \so(8)
\begin{align*}
 \so(8) \cong (\text{Ising})\times(\text{tricritical Ising}) \times
 (\text{3-state Potts})\times \su(3).
\end{align*}
In this decomposition, the $\su(3)$ part is absorbed as the holonomy,
but the statistical part $(\text{Ising})\times(\text{tricritical Ising})
\times (\text{3-state potts})$ remains unbroken and characterizes the
universal structures of Calabi-Yau compactifications, such as spacetime
supersymmetry and gauge group. We consider more about $\su(3)$ holonomy
and $\su(2)$ holonomy cases in the following subsections.

\subsubsection{\su(3) holonomy}
Let us first consider the \su(3) holonomy case. 
The prime example of this case is the Calabi-Yau compactification.
We expect there are $N=2$ superconformal symmetry and a spectral flow
operator. We explain how this symmetry can be seen from the
cascade of holonomies.

The class $\CFT(\su(3))$ is characterized by the coset
 \so(8)$/$\su(3), whose central charge is $c=2$. 
We denote this $c=2$ CFT as ${\cal X}$. One can
construct this ${\cal X}$ from minimal models by using the sequence
(\ref{seq-holonomy}) and Eqs.(\ref{coset1}),(\ref{coset2}).
We can define the characters of $\cal X$ by the set of equations
\begin{align}
\chi^{\so(8)}_{\lambda}
= \sum_{\Lambda}\chi^{
\so(8)/\su(3)}_{(\lambda,\Lambda)} \chi^{\su(3)}_{\Lambda},\quad 
\lambda=\basi,\vect,\spi,\cospi,\quad \Lambda=\basi,\fund,\cfund,
\end{align}
where, $\basi,\vect,\spi,\cospi,\fund,\cfund$ mean
 basic, vector, spinor, cospinor, fundamental, conjugate
fundamental representation respectively.
Then, from
Eqs.(\ref{so(8)/so(7)}),(\ref{so(7)/G2}) and (\ref{G2/su(3)}), the coset
characters can be written by using characters of minimal models. 
The results are collected as
  \begin{align}
   &\chi^{\so(8)/\su(3)}_{(\basi,\basi)}=\chi^{\rm min}_{(0,0,0)}+
                    \chi^{\rm min}_{(1/2,3/2,0)}+
                    \chi^{\rm min}_{(0,3/5,2/5)}+
                    \chi^{\rm min}_{(1/2,1/10,2/5)},\nn\\
   &\chi^{\so(8)/\su(3)}_{(\vect,\basi)}=\chi^{\rm min}_{(1/2,0,0)}+
                    \chi^{\rm min}_{(0,3/2,0)}+
                    \chi^{\rm min}_{(1/2,3/5,2/5)}+
                    \chi^{\rm min}_{(0,1/10,2/5)},\nn\\
   &\chi^{\so(8)/\su(3)}_{(\spi,\basi)}=
   \chi^{\so(8)/\su(3)}_{(\cospi,\basi)}=\chi^{\rm min}_{(1/16,7/16,0)}+
                    \chi^{\rm min}_{(1/16,3/80,2/5)},\nn\\
   &\chi^{\so(8)/\su(3)}_{(\basi,\fund)}=
   \chi^{\so(8)/\su(3)}_{(\basi,\cfund)}=
                    \chi^{\rm min}_{(0,0,2/3)}+
                    \chi^{\rm min}_{(1/2,3/2,2/3)}+
                    \chi^{\rm min}_{(0,3/5,1/15)}+
                    \chi^{\rm min}_{(1/2,1/10,1/15)},\nn\\
   &\chi^{\so(8)/\su(3)}_{(\vect,\fund)}=
   \chi^{\so(8)/\su(3)}_{(\vect,\cfund)}=
                    \chi^{\rm min}_{(1/2,0,2/3)}+
                    \chi^{\rm min}_{(0,3/2,2/3)}+
                    \chi^{\rm min}_{(1/2,3/5,1/15)}+
                    \chi^{\rm min}_{(0,1/10,1/15)},\nn\\
   &\chi^{\so(8)/\su(3)}_{(\spi,\fund)}=
   \chi^{\so(8)/\su(3)}_{(\cospi,\fund)}=
   \chi^{\so(8)/\su(3)}_{(\spi,\cfund)}=
   \chi^{\so(8)/\su(3)}_{(\cospi,\cfund)}=
                    \chi^{\rm min}_{(1/16,7/16,2/3)}+
                    \chi^{\rm min}_{(1/16,3/80,1/15)},
  \label{so(8)/su(3)min}
  \end{align}
 where $\chi^{\rm min}_{(a,b,c)}$ is the product of the characters of 
 minimal models
 \begin{align*}
  \chi^{\rm min}_{(a,b,c)}=\chi^{\isi}_a\chi^{\tri}_bC^{\pot}_c.
 \end{align*}
Also the symbol $(a,b,c)$ represents 
the set of conformal weights of each statistical model.
Since the ${\cal X}$ causes the gauge symmetry enhancement 
$\so(8)\to E_6$ 
(we are comparing $\CFT(\so(8))$ and $\CFT(\su(3))$),
the coset CFT $E_6/\so(8)$ 
is also identified with this ${\cal X}$. 
This fact can be seen from the sequence of gauge theory
(\ref{seq-gauge}) and Eqs.(\ref{coset1}),(\ref{coset2}).
We will also explain this fact from the point of view of characters later.

We can now obtain the characters of the coset CFT $E_6/\so(8)$ by using
the explicit forms of the characters shown in appendix \ref{appendix-WZW}.
When we define the coset characters
 $\chi^{E_6/\so(8)}_{(\Lambda,\lambda)}$'s by relations
\begin{align*}
 \chi^{E_6}_{\Lambda}=\sum_{\lambda}\chi^{E_6/
 \so(8)}_{(\Lambda,\lambda)}\chi^{\so(8)}_{\lambda},
\quad \Lambda=\basi,\fund,\cfund,\quad
 \lambda=\basi,\vect,\spi,\cospi,
\end{align*}
then we can obtain the results about characters
\begin{align}
 &\chi^{E_6/\so(8)}_{(\basi,\basi)}=
      \frac{\Theta_{0,6}}{\eta}\chi^{\so(2)}_{\basi}
     +\frac{\Theta_{6,6}}{\eta}\chi^{\so(2)}_{\vect},\nn\\
 &\chi^{E_6/\so(8)}_{(\basi,\vect)}=
      \frac{\Theta_{6,6}}{\eta}\chi^{\so(2)}_{\basi}
     +\frac{\Theta_{0,6}}{\eta}\chi^{\so(2)}_{\vect},\nn\\
 &\chi^{E_6/\so(8)}_{(\basi,\spi)}=\chi^{E_6/\so(8)}_{(\basi,\cospi)}=
      \frac{\Theta_{3,6}}{\eta}\chi^{\so(2)}_{\spi}
     +\frac{\Theta_{3,6}}{\eta}\chi^{\so(2)}_{\cospi},\nn\\
 &\chi^{E_6/\so(8)}_{(\fund,\basi)}=\chi^{E_6/\so(8)}_{(\cfund,\basi)}=
      \frac{\Theta_{4,6}}{\eta}\chi^{\so(2)}_{\basi}
     +\frac{\Theta_{2,6}}{\eta}\chi^{\so(2)}_{\vect},\nn\\
 &\chi^{E_6/\so(8)}_{(\fund,\vect)}=\chi^{E_6/\so(8)}_{(\cfund,\vect)}=
      \frac{\Theta_{2,6}}{\eta}\chi^{\so(2)}_{\basi}
     +\frac{\Theta_{4,6}}{\eta}\chi^{\so(2)}_{\vect},\nn\\
 &\chi^{E_6/\so(8)}_{(\fund,\spi)}=\chi^{E_6/\so(8)}_{(\cfund,\spi)}=
  \chi^{E_6/\so(8)}_{(\fund,\cospi)}=\chi^{E_6/\so(8)}_{(\cfund,\cospi)}=
      \frac{\Theta_{1,6}}{\eta}\chi^{\so(2)}_{\spi}
     +\frac{\Theta_{5,6}}{\eta}\chi^{\so(2)}_{\cospi}.
  \label{E6/so(8)}
\end{align}
We find the relation between the functions in Eqs.(\ref{so(8)/su(3)min})
and (\ref{E6/so(8)})
\begin{align}
 \chi^{E_6/\so(8)}_{(\Lambda,\lambda)}
       =\chi^{\so(8)/\su(3)}_{(\lambda,\Lambda)}, \label{3minimal-identity}
\end{align}
must be satisfied for each set $(\Lambda,\lambda)$
from the following reason. By using the explicit forms of
characters, the $E_8$ character can be decomposed by two \so(8)
characters as
\begin{align}
 \chi^{E_8}_{\basi}=\sum_{\lambda=\basi,\vect,\spi,\cospi}
     \chi^{\so(8)}_{\lambda}\chi^{\so(8)}_{\lambda}.
 \label{E8/so(8)}
\end{align}
Using the definition of $\chi^{\so(8)/\su(3)}$ and
 Eq.(\ref{E8/so(8)}), we can decompose $\chi^{E_8}_{\basi}$ in the
 following formula
\begin{align}
 \chi^{E_8}_{\basi}=\sum_{\Lambda=\basi,\fund,\cfund}
   \sum_{\lambda=\basi,\vect,\spi,\cospi}
     \chi^{\so(8)}_{\lambda}\chi^{\so(8)/\su(3)}_{(\lambda,\Lambda)}
     \chi^{\su(3)}_{\Lambda}.
  \label{E8/so(8)Xsu(3)}
\end{align}
On the other hand, by using the explicit forms of the characters,
the $E_8$ character can be decomposed by those of $E_6$ and $\su(3)$ as
\begin{align}
 \chi^{E_8}_{\basi}=\sum_{\Lambda=\basi,\fund,\cfund}
     \chi^{E_6}_{\Lambda} \chi^{\su(3)}_{\Lambda}.
  \label{E8/E6}
\end{align}
When we compare Eq.(\ref{E8/so(8)Xsu(3)}) and Eq.(\ref{E8/E6}), 
the set of relations is obtained
\begin{align}
 \chi^{E_6}_{\Lambda}=\sum_{\lambda=\basi,\vect,\spi,\cospi}
      \chi^{\so(8)}_{\lambda} \chi^{\so(8)/\su(3)}_{\lambda,\Lambda}.
  \label{E6/X}
\end{align}
By comparing the definition of $\chi^{E_6/\so(8)}$ and Eq.(\ref{E6/X}),
we can obtain the relation (\ref{3minimal-identity}).

We also checked the relation (\ref{3minimal-identity})
 for several order in $q$-expansion by Mathematica.
So we use the relation (\ref{3minimal-identity}) in this paper and 
introduce a notation 
$ \chi^{\cal X}_{(\Lambda,\lambda)}
       :=\chi^{E_6/\so(8)}_{(\Lambda,\lambda)}
       =\chi^{\so(8)/\su(3)}_{(\lambda,\Lambda)}$.

Looking at the forms in Eqs.(\ref{E6/so(8)}), 
we can show that the ${\cal X}$ is 
decomposed into \so(2) and u(1)${}_6$. This \so(2) corresponds to the
rotation of flat space $\Rb^2$ in transverse directions of Calabi-Yau
compactification CY${}_3\times \Rb^2$, and this u(1)${}_6$ is the
symmetry related to spacetime susy and gauge symmetry
enhancement $\so(10)\rightarrow E_6$. 
Actually, this $\uone$ symmetry can be identified with
the $\uone$ symmetry in the $c=9$, $N=2$ superconformal algebra. 
In order to see this, we will concentrate 
on the gauge symmetry enhancement \so(10)$\to E_6$. 
This phenomenon of the Calabi-Yau compactification 
can be realized as the relations about characters
\begin{align*} \chi^{E_6}_{\Lambda}=\sum_{\lambda}\frac{\Theta_{-4
n(\Lambda)+3n(\lambda),6}}{\eta} \chi^{\so(10)}_{\lambda},
\end{align*} 
where $n(\Lambda)$ and $n(\lambda)$ 
are functions respectively  
depending on representations $\Lambda$'s and $\lambda$'s
\begin{align*} n(\basi)=0,\ n(\fund)=n(\spi)=1,\
n(\cfund)=n(\cospi)=-1,\ n(\vect)=2.\ 
\end{align*}
Now let us write an arbitrary state in the Calabi-Yau CFT as
$|n,m\rangle\otimes |\text other\rangle$ where $|n,m\rangle$ is a
state in the module ``$m$'' of the $\uone_6$ theory.
The integer ``$m$'' appears 
as an index of the character $\Theta_{m,6}/\eta$.
The $|\text other\rangle$
is a state associated with other parts and 
has no contribution
to $\uone$ charge of the $N=2$ SCA. 
Only the part $|n,m\rangle$ has
the relevant $\uone$ charge. 
We can evaluate 
the $\uone$ charge $Q_m$ of this state as $Q_m=m/2 \mod 6$.
The $U(1)$ here serves as an R-symmetry of the 
$N=2$ theory and 
is used to construct $U(1)$ current of the 
$N=2$ algebra.
Also a spectral flow operator of the $N=2$ CFT 
has conformal dimension $3/8$ 
and is constructed 
by combining 
scaling operators of three statistical models.
A candidate of a spectral flow operator appears in the character
$\Theta_{3,6}/\eta$
, more precisely 
in the sector 
  \begin{align}
   &\chi^{\so(8)/\su(3)}_{(\spi,\basi)}=
   \chi^{\so(8)/\su(3)}_{(\cospi,\basi)}=\chi^{\rm min}_{(1/16,7/16,0)}+
                    \chi^{\rm min}_{(1/16,3/80,2/5)},\nn\\
    & = \frac{\Theta_{3,6}}{\eta}\chi^{\so(2)}_{\spi}
     +\frac{\Theta_{3,6}}{\eta}\chi^{\so(2)}_{\cospi}
    =\chi^{E_6/\so(8)}_{(\basi,\spi)}=\chi^{E_6/\so(8)}_{(\basi,\cospi)}
\label{cy}\,.
  \end{align}
This operator belongs to a sector 
with a $U(1)$ charge $Q=3/2$. 
It implies that the state is related with 
a $3$-form of the $CY_3$.
Also the lowest term in this character is $q^{3/8-1/24}$.
This represents a primary state with conformal weight $3/8$ and its
charge is $ 3/2$. It is the same as the spectral flow
operator $\Sigma$ has. 
We shall look at this more precisely. 
It is realized as a combination of states 
with $(h^{\isi},h^{\tri},h^{\pot})=(1/16,7/16,0),(1/16,3/80,2/5)$. 
The total weight of these states turns out to be $1/2$. 
When we recall the identity
$\chi^{\isi}\times \chi^{\tri}\times
\chi^{\pot}=
\chi^{\SO(2)}\times \chi^{U(1)}$ in the above Eqs.(\ref{cy}), 
we can obtain a spin operator $\Sigma$ with $h=3/8(=1/2-1/8)$ of the 
$\SU(3)$ holonomy model 
by subtracting contributions of a spin operator of $\SO(2)$ 
with weight $1/8$.
This operator $\Sigma$ 
is nothing but a holomorphic $3$-form of the $CY_3$ and 
confirms the validity of our discussions. 
(But we only look at the chiral part of the theory here).

It is remarkable that we can realize 
$N=2$ CFT associated with $CY_3$ starting from $\SO(8)$ theory 
by using three statistical models in $2$ dimension.

\subsubsection{\su(2) holonomy}
Let us also consider the \su(2) holonomy case in the same way as the \su(3)
case. The prime example of this case is the K3 compactification. In
the \su(2) holonomy case, the coset $\so(8)/\su(2)$ is
essential to explain spacetime 
supersymmetry and the gauge symmetry enhancement.
We denote $\so(8)/\su(2)$ as ${\cal Y}$ and study how this 
determines the spacetime susy and gauge symmetry.

The characters of ${\cal Y}$ are defined by the branching relation
\begin{align*}
 &\chi^{\so(8)}_{\lambda}=\sum_{\Lambda=
\basi,\fund} \chi^{\cal Y}_{(\Lambda,\lambda)}\chi^{\su(2)}_{\Lambda}.
\end{align*}
By using $\so(8)/\su(3)\cong {\cal X}$ in
Eqs.(\ref{so(8)/su(3)min}), $\su(3)/\su(2)\cong \uone_3$ in
Eqs.(\ref{su(3)/su(2)}), and the explicit forms of the $\chi^{\cal
X}_{(\Lambda,\lambda)}$'s in Eqs.(\ref{E6/so(8)}), the characters of ${\cal
Y}$ can be written as
\begin{align*}
 &\chi^{\cal Y}_{(\basi,\basi)}
         =\chi^{\so(4)}_{\basi}\chi^{\su(2)}_{\basi},
 \qquad\chi^{\cal Y}_{(\basi,\vect)}
         =\chi^{\so(4)}_{\vect}\chi^{\su(2)}_{\basi},\\
 &\chi^{\cal Y}_{(\fund,\basi)}
         =\chi^{\so(4)}_{\vect}\chi^{\su(2)}_{\fund},
 \qquad\chi^{\cal Y}_{(\fund,\vect)}
         =\chi^{\so(4)}_{\basi}\chi^{\su(2)}_{\fund},\\
 &\chi^{\cal Y}_{(\basi,\spi)}=\chi^{\cal Y}_{(\basi,\cospi)}
         =\chi^{\so(4)}_{\spi}\chi^{\su(2)}_{\fund},
 \qquad\chi^{\cal Y}_{(\fund,\spi)}=\chi^{\cal Y}_{(\fund,\cospi)}
         =\chi^{\so(4)}_{\spi}\chi^{\su(2)}_{\basi}.
\end{align*}
These relations show that ${\cal Y}$ can be decomposed into \so(4) and
\su(2). This \so(4) in ${\cal Y}$ is the rotation of flat $\Rb^4$ in
$\Rb^4\times$K3.  On the other hand, this 
$\su(2)$ in ${\cal Y}$ is the
key symmetry for the spacetime supersymmetry and the gauge symmetry
enhancement. Actually, the K3 CFT has $c=6$, N=4 superconformal symmetry
and the $\su(2)$ in ${\cal Y}$ is identified with $R$-symmetry $\su(2)$ 
in the $c=6$, N=4 superconformal algebra.

The gauge symmetry is enhanced from $\so(12)$ to $E_7$
with this $\su(2)$ symmetry in ${\cal Y}$. 
This phenomenon can be explained by branching rules
\begin{align}
 &\chi^{E_7}_{\basi}=\chi^{\SU(2)}_{\basi}\chi^{\SO(12)}_{\basi}
 +\chi^{\SU(2)}_{\fund}\chi^{\SO(12)}_{\spi}\,,\non
 &\chi^{E_7}_{\fund}=\chi^{\SU(2)}_{\basi}\chi^{\SO(12)}_{\cospi}
 +\chi^{\SU(2)}_{\fund}\chi^{\SO(12)}_{\vect}\,.\nom
\end{align}

The spectral flow operators are the primary states in the fundamental
representation of $\su(2)$ in ${\cal Y}$. The conformal dimension of
these states are both $1/4$. This is the same property as spectral
flow operators.

\subsubsection{Comments on spacetime supersymmetry}
Let us comment about the spacetime supersymmetry from the viewpoint
of characters. In the flat case, the key ingredient for this susy
is the Jacobi's abstruse identity
\begin{align*}
 \theta_3^4-\theta_4^4-\theta_2^4=0.
\end{align*}
This can be rewritten by \so(8) characters in the formula
\begin{align*}
 \chi^{\so(8)}_{\vect}-\chi^{\so(8)}_{\spi}=0.
\end{align*}
{} From this Jacobi's abstruse identity, we can propose the key identities
for the spacetime supersymmetries in compactifications on special
holonomy manifolds
\begin{align}
 \chi^{\so(8)/\Ghol}_{\vect,\Lambda}-\chi^{\so(8)/\Ghol}_{\spi,\Lambda}=0,
\label{super-identity}
\end{align}
where $\Ghol$'s are the holonomy groups.
Also $\Lambda$ is the representation of
$\Ghol$. We assume this identity is satisfied only for the cases
$\chi^{\so(8)/\Ghol}_{\vect,\lambda}\ne 0$ and
$\chi^{\so(8)/\Ghol}_{\spi,\lambda}\ne0$.
An evidence of our proposal is given by the following
branching relation
\begin{align*}
0= \chi^{\so(8)}_{\vect}-\chi^{\so(8)}_{\spi}
=\sum_{\Lambda} (\chi^{\so(8)/\Ghol}_{\vect,\Lambda}
                 -\chi^{\so(8)/\Ghol}_{\spi,\Lambda})
   \chi^{\Ghol}_{\Lambda}.
\end{align*}

If we put the identity (\ref{super-identity}), we can show that
the partition functions vanish in orbifold cases.
In order to explain this, let us consider 
the orbifold group $\Gamma \subset\Ghol\subset\so(8)$.
The character of the $(g_1,g_2)$-sector $(g_1,g_2 \in \Gamma )$ is defined as
\begin{align*}
 \chi^{\so(8)}_{\lambda,(g_1,g_2)}
=\Tr_{\lambda, g_1 {\rm twisted}}[g_2 q^{L_0-c/24}].
\end{align*}
This can be decomposed as
\begin{align*}
&\chi^{\so(8)}_{\lambda,(g_1,g_2)}
=\sum_{\Lambda}\chi^{\so(8)/\Ghol}_{\lambda ,\Lambda}
   \chi^{\Ghol}_{\Lambda,(g_1,g_2)}.
\end{align*}
Note that $\chi^{\so(8)/\Ghol}_{\lambda ,\Lambda}$ is independent of
$(g_1,g_2)$ because $g_1$ and $g_2$ are elements of $\Gamma\subset \Ghol$. 
On the other hand, 
$\chi^{\Ghol}_{\Lambda,(g_1,g_2)}$ is defined in the same way as the
$\so(8)$ case.
By using these characters, the partition function of left-moving fermions
in $(g_1,g_2)$-sector can be written as
\begin{align*}
 Z^{(F)}_{g_1,g_2}
   = \chi^{\so(8)}_{\vect,(g_1,g_2)}-\chi^{\so(8)}_{\spi,(g_1,g_2)}
   =\sum_{\Lambda} (\chi^{\so(8)/\Ghol}_{\vect,\Lambda}
                       -\chi^{\so(8)/\Ghol}_{\spi,\Lambda})
        \chi^{\Ghol}_{\Lambda,(g_1,g_2)}.
\end{align*}
This $Z^{(F)}_{g_1,g_2}$ becomes $0$ when we use the identities
(\ref{super-identity}).

Let us see the explicit forms of these identities for each case of
holonomies $G_2,\su(3),\su(2)$.  First, we consider the explicit form of
the identities of the $G_2$ case. The branching relation of the coset
model $\so(8)/G_2$ can be written as
\begin{align*}
 &\chi^{\so(8)}_{\basi}=
 (\chi^{\isi}_{0}\chi^{\tri}_{0}+\chi^{\isi}_{1/2}\chi^{\tri}_{3/2})
                      \chi^{G_2}_{\basi}+
 (\chi^{\isi}_{0}\chi^{\tri}_{3/5}+\chi^{\isi}_{1/2}\chi^{\tri}_{1/10})
                      \chi^{G_2}_{\fund},\\
 &\chi^{\so(8)}_{\vect}=
 (\chi^{\isi}_{1/2}\chi^{\tri}_{0}+\chi^{\isi}_{0}\chi^{\tri}_{3/2})
                      \chi^{G_2}_{\basi}+
 (\chi^{\isi}_{1/2}\chi^{\tri}_{3/5}+\chi^{\isi}_{0}\chi^{\tri}_{1/10})
                      \chi^{G_2}_{\fund},\\
 &\chi^{\so(8)}_{\spi}=\chi^{\so(8)}_{\cospi}=
 \chi^{\isi}_{1/16}\chi^{\tri}_{7/16}
                      \chi^{G_2}_{\basi}+
 \chi^{\isi}_{1/16}\chi^{\tri}_{3/80}
                      \chi^{G_2}_{\fund}.
\end{align*}
The explicit susy identities in the $G_2$ holonomy case is as follows. From 
 $\chi^{\so(8)/G_2}_{(\vect,\basi)}-\chi^{\so(8)/G_2}_{(\spi,\basi)}=0$ ,
we obtain
\begin{align}
 &\chi^{\isi}_{1/2}\chi^{\tri}_{0}+\chi^{\isi}_{0}\chi^{\tri}_{3/2}
    -\chi^{\isi}_{1/16}\chi^{\tri}_{7/16}=0.
 \label{susyG2-1}
\end{align}
 From 
 $\chi^{\so(8)/G_2}_{(\vect,\fund)}-\chi^{\so(8)/G_2}_{(\spi,\fund)}=0$ ,
we obtain
\begin{align}
 &\chi^{\isi}_{1/2}\chi^{\tri}_{3/5}+\chi^{\isi}_{0}\chi^{\tri}_{1/10}
 -\chi^{\isi}_{1/16}\chi^{\tri}_{3/80}=0.
 \label{susyG2-2}
\end{align}
These formulas are the same as the ones recently obtained in 
\cite{Eguchi:2001xa}.

Next, let us go to the susy identities in the $\su(3)$ holonomy case.
The explicit form using the characters in (\ref{so(8)/su(3)min})
becomes
\begin{align}
& \chi^{\rm min}_{(1/2,0,0)}+\chi^{\rm min}_{(0,3/2,0)}+
 \chi^{\rm min}_{(1/2,3/5,2/5)}+\chi^{\rm min}_{(0,1/10,2/5)}
 -\chi^{\rm min}_{(1/16,7/16,0)}-\chi^{\rm min}_{(1/16,3/80,2/5)}=0,\nn\\
&\chi^{\rm min}_{(1/2,0,2/3)}+ \chi^{\rm min}_{(0,3/2,2/3)}+
  \chi^{\rm min}_{(1/2,3/5,1/15)}+\chi^{\rm min}_{(0,1/10,1/15)}
  -\chi^{\rm min}_{(1/16,7/16,2/3)}-\chi^{\rm min}_{(1/16,3/80,1/15)}=0.
\label{susySU(3)-1}
\end{align}
Since a Calabi-Yau compactification is a special case of $G_2$
compactifications, one may guess that the identities (\ref{susySU(3)-1})
can be {\em derived} from the identities (\ref{susyG2-1}) and
(\ref{susyG2-2}).  Actually, the following formulas show this
guess is true
\begin{align*}
& \chi^{\rm min}_{(1/2,0,0)}+\chi^{\rm min}_{(0,3/2,0)}+
 \chi^{\rm min}_{(1/2,3/5,2/5)}+\chi^{\rm min}_{(0,1/10,2/5)}
 -\chi^{\rm min}_{(1/16,7/16,0)}-\chi^{\rm min}_{(1/16,3/80,2/5)}\\
&\qquad=(\chi^{\so(8)/G_2}_{(\vect,\basi)}-\chi^{\so(8)/G_2}_{(\spi,\basi)})
C^{\pot}_{0}+
(\chi^{\so(8)/G_2}_{(\vect,\fund)}-\chi^{\so(8)/G_2}_{(\spi,\fund)})
C^{\pot}_{5/2},\\
& \chi^{\rm min}_{(1/2,0,2/3)}+\chi^{\rm min}_{(0,3/2,2/3)}+
 \chi^{\rm min}_{(1/2,3/5,1/15)}+\chi^{\rm min}_{(0,1/10,1/15)}
 -\chi^{\rm min}_{(1/16,7/16,2/3)}-\chi^{\rm min}_{(1/16,3/80,1/15)}\\
&\qquad=(\chi^{\so(8)/G_2}_{(\vect,\basi)}-\chi^{\so(8)/G_2}_{(\spi,\basi)})
C^{\pot}_{2/3}+
(\chi^{\so(8)/G_2}_{(\vect,\fund)}-\chi^{\so(8)/G_2}_{(\spi,\fund)})
C^{\pot}_{1/15}.
\end{align*}
Besides the expression (\ref{susySU(3)-1}) of the susy identities, we
can also write the explicit susy identities using the form of
(\ref{E6/so(8)}).  These identites reduce to
\begin{align}
 &\Theta_{6,6}\Theta_{0,2}
     +\Theta_{0,6}\Theta_{2,2}-
  2\Theta_{3,6}\Theta_{1,2}=0,\nn\\
 &\Theta_{2,6}\Theta_{0,2}
     +\Theta_{4,6}\Theta_{2,2}-
  \Theta_{1,6}\Theta_{1,2}-
  \Theta_{5,6}\Theta_{1,2}=0.
\label{susySU(3)-2}
\end{align}
These are the same identites obtained in \cite{Eguchi:1989vr}
and \cite{Kutasov:1991pv}. If we use the identity
(\ref{3minimal-identity}), (\ref{susySU(3)-1}) and (\ref{susySU(3)-2})
are equivalent.

Finally, let us see the susy identities in the $\su(2)$ holonomy case.
The explicit form of the identities are given by
\begin{align*}
&\chi^{\so(4)}_{\vect}\chi^{\su(2)}_{\basi}
         -\chi^{\so(4)}_{\spi}\chi^{\su(2)}_{\fund}=0,\\
&\chi^{\so(4)}_{\basi}\chi^{\su(2)}_{\fund}
       -\chi^{\so(4)}_{\spi}\chi^{\su(2)}_{\basi}=0.
\end{align*}
These identities reduce to the ones obtained in
\cite{Bilal:1987uh}.

\section{Orbifold}

In this section we investigate
$G_2$ and $Spin(7)$ manifolds 
realized as orbifolds.
These models have been discussed by
Joyce\cite{Joyce1,Joyce2,Joyce3} as 
concrete examples of compact 
manifolds with exceptional holonomies in mathematical 
contexts.
We review his constructions in 
subsection 5.1. 
In subsection 5.2, we 
elaborate 
toroidal partition functions of heterotic strings 
on these orbifolds and study their modular 
properties. 
In subsection 5.3, we 
show our results about massless spectra 
of effective theories in our these heterotic models.

\subsection{Examples of Special Holonomy Manifolds}

In this subsection we study some of the examples constructed by
Joyce\cite{Joyce1,Joyce2,Joyce3}.
A basic example of a compact seven manifold $M^{(7)}$ with $G_2$ holonomy is 
realized as a toroidal orbifold. 
Let $(x_1, x_2, \cdots , x_7)$ be a set of coordinates 
of $T^7$ which is a product of seven circles of 
the radius $R$.
The $M^{(7)}$ is defined as the 
desingularization of the $T^7$ modded out by 
$\Gamma\cong {\bz}^3_2$ group with generators
\ba
&&T^7\ni(x_1,x_2,x_3,x_4,x_5,x_6,x_7)\,,\non
&&\Gamma\left\{
\begin{array}{cccccccccc}
\alpha ; & ( & -x_1, & -x_2, & -x_3, & -x_4, & x_5, & x_6, &  x_7 & )\\
\beta ;& ( & -x_1, & \dis \frac{1}{2}-x_2, 
& x_3, & x_4, & -x_5, & -x_6, 
&  x_7 & )\\
\gamma ; & ( & \dis\frac{1}{2}-x_1, 
& x_2, & \dis\frac{1}{2}-x_3, & x_4, & -x_5, & x_6, & -x_7 & )
\end{array}
\right.
\,,\nom
\ea
where the generators of the ${\bz}_2$'s are denoted by $\alpha$,
$\beta$, and $\gamma$. 
One can verify that 
$\alpha^2=\beta^2=\gamma^2=1$ 
and $\alpha$, $\beta$, $\gamma$ commute one another.
Then discrete group $\Gamma$ is isomorphic to ${\bz}^3_2$. 
Also $1/2$ means a shift $\dis \frac{1}{2}\times 2\pi R$ around the 
circle in the case that each $x_i$ of $T^7$ has period $2\pi R$.
Then this holonomies preserve the flat $G_2$ structure on $T^7$
given by a $\Phi$
\ba
&&\Phi =d{\bf x}_{136}+d{\bf x}_{145}+d{\bf x}_{235}-d{\bf x}_{127}
-d{\bf x}_{246}-d{\bf x}_{347}-d{\bf x}_{567}\,,\non
&&d{\bf x}_{ijk}:=dx_i\wedge dx_j\wedge dx_k\,.\nom
\ea
Next we review the cohomology classes on $M^{(7)}$. 
After this projection there remain a zero-form, one $7$-form, 
seven $3$-forms and seven $4$-forms
of $T^7$. But none of the two-forms are invariant under the action 
of the discrete group $\Gamma$.
The elements $\beta\gamma$, $\gamma\alpha$, $\alpha\beta$ and 
$\alpha\beta\gamma$ have no fixed points on $T^7$. The fixed points of 
$\alpha$ in $T^7$ are $16$ $T^3$'s and the group 
$\langle \beta ,\gamma\rangle$ acts freely on these 16 sets. 
It leaves us with $4$ invariant combinations on the quotient 
$T^7/\Gamma$. Similarly one can see that 
the fixed $T^3$'s for each $\beta$,$\gamma$ are $4$ copies of $T^3$. 
The local form of the singularities at the fixed $T^3$'s 
is ${\bf R}^4/{\bz}_2 \times T^3$ and resolving each of these 
yields one $2$-form and three $3$-forms. 
Since there are $12$ fixed tori on $M^{(7)}$, 
one obtains Betti numbers after resolution
by recalling $b_2(T^7/\Gamma)=0$, $b_3(T^7/\Gamma)=7$
\ba
b_2(M^{(7)})=b_2(T^7/\Gamma)+12\cdot 1=12\,,\,\,
b_3(M^{(7)})=b_3(T^7/\Gamma)+12\cdot 3=43\,.\,\,\nom
\ea
Now we are able to write down all Betti numbers of this $G_2$ orbifold $M^{(7)}$
\ba
&& b_0=b_7=1\,,\,\,b_1=b_6=0\,,\,\,\non
&&b_2=b_5=12\,,\,\,b_3=b_4=43\,.\nom
\ea
This is a compact, simply-connected seven manifold with holonomy
$G_2$.
The moduli space has dimension $43$ and 
the associated $CFT$ counterpart is 
a $b_2+b_3=55$ dimensional space.

Next we shall explain a simple example of a 
compact $8$ manifold $M^{(8)}$ with holonomy $Spin(7)$ constructed by
Joyce\cite{Joyce3}. This example proceeds similarly to the $G_2$ case. 
Let $(x_1, x_2, \cdots , x_7,x_8)$ be a set of coordinates 
of $T^8$ which is a product of eight circles of 
the radius $R$.
The $M^{(8)}$ is constructed as the 
desingularization of the $T^8$ divided by the discreet group
$\Gamma\cong {\bz}^4_2$ with generators
\ba
&&T^8 \ni(x_1,x_2,x_3,x_4,x_5,x_6,x_7,x_8)\,,\non
&&\Gamma\left\{
\begin{array}{ccccccccccc}
\alpha ; & ( & -x_1, & -x_2, & -x_3, & -x_4, & x_5, & x_6, &  x_7, 
& x_8 & )\\
\beta ;& ( & x_1, & x_2, & x_3, & x_4, & -x_5, & -x_6, &  -x_7, 
& -x_8 & )\\
\gamma ; & ( & \dis\frac{1}{2}-x_1, & \dis\frac{1}{2}-x_2, & x_3, & x_4,
 & \dis\frac{1}{2}-x_5, 
& \dis\frac{1}{2}-x_6, & 
\dis x_7, & x_8 & )\\
\delta ; & ( & \dis -x_1, & x_2, & \dis \frac{1}{2}-x_3, & x_4,
 & \dis -x_5, 
& \dis x_6, & 
\dis \frac{1}{2}-x_7, & x_8 & )
\end{array}
\right.
\,\,.\nom
\ea
It is easy to see that
$\alpha^2=\beta^2=\gamma^2=\delta^2=1$ and 
$\alpha,\beta,\gamma ,\delta$ all commute one another.
Then the $\Gamma\cong {\bz}^4_2$ is a 
group of automorphisms of $T^8$ preserving 
the flat $Spin(7)$ structure given by a Cayley $4$ form 
$\Phi$
\ba
&&\Phi =d{\bf x}_{1234}+d{\bf x}_{1256}+d{\bf x}_{1278}+
d{\bf x}_{1357}-d{\bf x}_{1368}\non
&&\qquad -d{\bf x}_{1458}-d{\bf x}_{1467}
-d{\bf x}_{2358}-d{\bf x}_{2367}-d{\bf x}_{2457}\non
&&\qquad +d{\bf x}_{2468}+d{\bf x}_{3456}+
d{\bf x}_{3478}+d{\bf x}_{5678}\,,\non
&&d{\bf x}_{ijk\ell}=dx_i\wedge dx_j\wedge dx_k\wedge dx_{\ell}\,.\nom
\ea
Each of the fixed points of $\alpha$,$\beta$,$\gamma$,$\delta$ 
are $16$ copies of $T^4$. 
Also $\beta$ acts trivially on the set of the $16$ $T^4$ 
fixed by the $\alpha$-action and 
$\langle\gamma ,\delta\rangle$ acts freely on these $T^4$. 
It leaves us with $4$ invariant combinations $T^4/\{\pm 1\}$ on the 
quotient $T^8/\Gamma$ from the $\alpha$-fixed points.
We summarize
similar properties about fixed $T^4$'s by other generators 
in table \ref{orb1},\ref{orb2}
\begin{table}
\ba
\begin{array}{|c|c|c|c|c|c|}\hline
 & \mbox{fixed points} 
& \alpha-\mbox{action} &\beta-\mbox{action} &\gamma-\mbox{action} 
&\delta-\mbox{action} \\\hline
\alpha-\mbox{fixed points} & 16\,\, T^4 
& \ast & \mbox{trivial}& \mbox{free}& \mbox{free}\\\hline
\beta-\mbox{fixed points} & 16\,\, T^4
& \mbox{trivial} & \ast & \mbox{free}& \mbox{free}\\\hline
\gamma-\mbox{fixed points} & 16 \,\, T^4 
& \mbox{free} & \mbox{free}& \ast & \mbox{free}\\\hline
\delta-\mbox{fixed points} & 16 \,\, T^4 
& \mbox{free} & \mbox{free}& \mbox{free}& \ast \\\hline
\end{array}\nom
\ea
\caption{Fixed-point sets by $\alpha$, $\beta$, $\gamma$, $\delta$ and
actions of these generators on them.}
\label{orb1}
\end{table}
\begin{table}
\ba
\begin{array}{|c|c|}\hline
 & \mbox{fixed points} 
\\\hline
\alpha-\mbox{fixed points} & 4\,\, T^4/\{\pm 1\} 
\\\hline
\beta-\mbox{fixed points} & 4\,\, T^4/\{\pm 1\} 
\\\hline
\gamma-\mbox{fixed points} & 2 \,\, T^4
\\\hline
\delta-\mbox{fixed points} & 2 \,\, T^4
\\\hline
\end{array}\nom
\ea
\caption{Fixed-point sets divided by actions of other generators.}
\label{orb2}
\end{table}
The two sets of $\alpha$-fixed points and $\beta$-fixed points
intersect in $64$ points.

As is the $G_2$ case, the Betti numbers $b_{\ell}(T^8/\Gamma)$ 
are the dimension of the $\Gamma$-invariant subspaces of 
differential forms on $T^8$. After the projection 
there are no nonzero
$\Gamma$-invariant $1$-, $2$-, and $3$-forms. 
But one can show that there are four self-dual $4$-forms and 
four anti self-dual $4$-forms. 
Thus the Betti numbers of $T^8/\Gamma$ are 
written down as
\ba
&&b_1(T^8/\Gamma)=b_2(T^8/\Gamma)=b_3(T^8/\Gamma)=0\,,\non
&&b_4(T^8/\Gamma)=14\,,\,\,
b_4^{+}(T^8/\Gamma)=7 \,,\,\,b_4^{-}(T^8/\Gamma)=7\,.\nom
\ea

Next we calculate the Betti numbers of $M^{(8)}$. 
When one resolves each of the $4$ fixed $T^4/\{\pm 1\}$ by $\alpha$-action 
and $4$ $T^4/\{\pm 1\}$ fixed by $\beta$ in $T^8/\Gamma$, 
the $b_3$ is unchanged but $1$ is added to the $b_2$. 
Also $3$ is added to each of $b_4^{\pm }$. 
For each of the $2$ $T^4$ fixed by $\gamma$-action and $2$ fixed $T^4$ 
by $\delta$ in $T^8/\Gamma$, there are contributions 
$1$ to $b_2$ and $3$ to each $b_4^{\pm}$. 
When we 
resolve each of the $64$ points in the intersection of the 
$4$ $\alpha$-fixed sets 
$T^4/\{\pm 1\}$ and the $4$ $\beta$-fixed $T^4/\{\pm 1\}$'s, 
this operation  
does not change $b_2$, $b_3$ and $b_4^{-}$ but 
adds $1$ to $b_4^{+}$. 
By collecting all the contributions, we obtain 
the Betti numbers of the $b_{\ell}(M^{(8)})$
\ba
&&b_0=b_8=1\,,\,\,b_1=b_7=0\,,\non
&&b_2=b_6=12\,,\,b_3=b_5=16\,,\,b_4=150\,,\,\non
&&b_4^{+}=107\,,\,b_4^{-}=43\,.\nom
\ea
In this model, the moduli space of holonomy $Spin(7)$ metrics on $M^{(8)}$ is
a smooth manifold of dimension $1+b_4^{-}=44$.

\subsection{Modular invariant partition function}
In this subsection, we write down the partition functions of the
orbifold string models explained in the previous subsection.  In this
paper, we work in light-cone gauge.

\subsubsection{$G_2$ holonomy manifold case}
First, we consider the $G_2$ compactification.  In this case, we set
$x^1,\dots,x^7$ to be the coordinates of the $G_2$ manifold, and $x^8$
to be the transverse direction of the flat spacetime.

In our model, only one of the two $E_8$ has 
information about the holonomy group, and the other
$E_8$ does not have any contribution of 
holonomy group of the internal manifold. 
We denote the $E_8$ including the holonomy group as
$E_8^{(1)}$ and the other as $E_8^{(2)}$. We describe $E_8^{(1)}$ by 
16 free fermions $\lt^1,\dots,\lt^{16}$. Among them,
$\lt^{1},\dots,\lt^{7}$ are orbifolded in the same way as 
the left-moving fermions $\psi^1,\dots,\psi^7$, 
and others are not orbifolded. Therefore, the $\soh(9)$
of $\lt^8,\dots,\lt^{16}$ is manifestly realized.

The orbifold partition 
functions $Z(\tau ,\bar{\tau})$ generally have the following form
\begin{align}
 &Z(\tau,\taub)=\frac{1}{|\Gamma|}\sum_{g_1,g_2\in \Gamma}
Z_{g_1,g_2}(\tau,\taub),\qquad
 Z_{g_1,g_2}(\tau,\taub)=\Tr_{g_1 \text{twisted sector}}
[g_2\;q^{L_0-c/24}\bar q^{\bar L_0-\bar{c}/24}],
\label{total-partition-function}
\end{align}
where $q=\exp(2\pi i\tau)$. 
Also $\tau$ is the modulus of the 
toroidal worldsheet.
The subscript $g_1$ represents 
twisted boundary condition along the 
$\sigma_1$ (spatial) direction on the 
worldsheet. On the other side, the $g_2$ 
expresses the boundary condition 
along the temporal direction on the 
worldsheet. The $(L_0-c/24,\bar{L}_0-\bar{c}/24)$ 
is the set of 
Hamiltonians on the left- and right-moving parts
in our heterotic string with $(c,\bar{c})=(12,24)$.
To be modular invariant, the following
modular properties should be satisfied
\begin{align}
 Z_{g_1,g_2}(-1/\tau)=Z_{g_2,g_1^{-1}}(\tau),\qquad
 Z_{g_1,g_2}(\tau+1)=Z_{g_1,g_1g_2}(\tau). \label{modular-invariance}
\end{align}
In our case, each $Z_{g_1,g_2}$ can be decomposed into 
several blocks
and can be written as a product of them
\begin{align}
  &Z_{g_1,g_2}(\tau,\taub)=
Z^{(\text{flat boson})}(\tau,\taub)Z^{(B)}_{g_1,g_2}(\tau,\taub)
\times Z^{(F)}_{g_1,g_2}(\tau)
\times \overline{(\chi^{E_8}_{g_1,g_2}(\tau)\chi^{E_8}_{\basi}(\tau))}.
\label{form-of-partition-function}
\end{align}
In this formula, $Z^{(\text{flat boson})}$ is the partition function of a
single boson $x^8$ ; $Z^{(\text{flat boson})}(\tau,\taub)=({\rm
Im}\tau)^{-1/2}|\eta(\tau)|^{-2}$. $Z^{(B)}_{g_1,g_2}$ is the 
partition function of the bosons $x^1,\dots,x^7$ in the  $g_1$-twisted sector
with $g_2$-insertion. This part describes 
the internal $G_2$ manifold. 
The $Z^{(B)}_{g_1,g_2}$'s themselves
satisfy the modular properties (\ref{modular-invariance}).
Also, $Z^{(F)}_{g_1,g_2}(\tau)$ is the character of the left-moving fermions
$\psi^1,\dots,\psi^7$ of $g_1$-twisted sector with $g_2$-insertion.
As a result of spacetime supersymmetry, 
each of $Z^{(F)}_{g_1,g_2}(\tau)$'s vanishes.
Next we consider structures on the right-moving part.
$\chi^{E_8}_{g_1,g_2}(\tau)$ is
the character of $E_8^{(1)}$ in the 
$g_1$-twisted sector with $g_2$-insertion.
$\chi^{E_8}_{\basi}(\tau)$ is the $E_8^{(2)}$ character defined as
\begin{align*}
 \chi^{E_8}_{\basi}(\tau)=\frac{1}{2\eta(\tau)^{8}}
\{\theta_3(\tau)^8+\theta_4(\tau)^8+\theta_2(\tau)^8\}.
\end{align*}
The explicit formulae of $Z^{(B)}_{g_1,g_2}$, $Z^{(F)}_{g_1,g_2}(\tau)$
and $\chi^{E_8}_{g_1,g_2}(\tau)$ are concretely calculated in our model.

First, we consider the boson sector $Z^{(B)}_{g_1,g_2}$.
Our orbifold group does not mix the coordinates one another, so we can
concentrate on each $x_i$ separately.
We have only to think the following 4 types of twistings
\begin{align*}
 \0: x\to x,\quad\1:x\to x+\frac12,\quad\2:x\to-x,\quad\3:x\to\frac12-x
\end{align*}
\begin{itemize}
\item[] The \1 -twisted sector differs from untwisted sector by zero-modes.
In \1 -twisted sector, the winding number becomes a half integer.
\item[] The \2 -twisted 
sector expresses an anti-periodic boson and it has half integral
modes.
\item[] The \3 -twisted 
sector is the same as \2 -twisted sector: when we define
$y=\frac14-x$, then \3 is rewritten as $y\to -y$.
\item[] The \1 -operator 
insertion contributes $(-1)^n$ where $n$ is the momentum.
\item[] The \2 -operator 
insertion is represented on oscillators $\alpha_{n}\to -\alpha_{n}$.
For zero-modes, only the zero momentum and zero winding part survives.
\item[] The \3 -operator 
insertion is the same as \2 -operator insertion.
\end{itemize}
As a result, we obtain the following partition function
 of a single boson $Z^{(B1)}_{ab}\ (a,b=\0,\1,\2,\3) $.
\begin{align}
&Z^{(B1)}_{\0\0}=|\eta(\tau)|^{-2}\sum_{n,w\in \Zb}
q^{\frac12 \left(\frac nR+\frac{Rw}{2}\right)^2}
\bar q^{\frac12 \left(\frac nR-\frac{Rw}{2}\right)^2},\nn\\
&Z^{(B1)}_{\0\1}=|\eta(\tau)|^{-2}\sum_{n,w\in \Zb}(-1)^n
q^{\frac12 \left(\frac nR+\frac{Rw}{2}\right)^2}
\bar q^{\frac12 \left(\frac nR-\frac{Rw}{2}\right)^2},\nn\\
&Z^{(B1)}_{\1\0}=|\eta(\tau)|^{-2}\sum_{n,w\in \Zb}
q^{\frac12 \left(\frac nR+\frac{R(w+1/2)}{2}\right)^2}
\bar q^{\frac12 \left(\frac nR-\frac{R(w+1/2)}{2}\right)^2},\nn\\
&Z^{(B1)}_{\1\1}=|\eta(\tau)|^{-2}\sum_{n,w\in \Zb}(-1)^n
q^{\frac12 \left(\frac nR+\frac{R(w+1/2)}{2}\right)^2}
\bar q^{\frac12 \left(\frac nR-\frac{R(w+1/2)}{2}\right)^2},\nn\\
&Z^{(B1)}_{\0\2}=Z^{(B1)}_{\0\3}
=\left|q^{-\frac1{24}}\prod_{n=1}^{\infty}(1+q^n)^{-1} \right|^2
=2\left|\frac{\eta(\tau)}{\theta_2(\tau)}\right|,\nn\\
&Z^{(B1)}_{\2\0}=Z^{(B1)}_{\3\0}
=2\left|q^{\frac1{48}}\prod_{n=1}^{\infty}(1-q^{n-\frac12})^{-1}
 \right|^2
=2\left|\frac{\eta(\tau)}{\theta_4(\tau)}\right|,\nn\\
&Z^{(B1)}_{\2\2}=Z^{(B1)}_{\3\3}
=2\left|q^{\frac1{48}}\prod_{n=1}^{\infty}(1+q^{n-\frac12})^{-1} \right|^2
=2\left|\frac{\eta(\tau)}{\theta_3(\tau)}\right|,\nn\\
&Z^{(B1)}_{\1\2}=Z^{(B1)}_{\1\3}=Z^{(B1)}_{\2\1}
=Z^{(B1)}_{\3\1}=Z^{(B1)}_{\2\3}=Z^{(B1)}_{\3\2}=0.\label{boson-character}
\end{align}
The $\eta$ is the Dedekind's eta function and 
$\theta_i$'s ($i=2,3,4$) represent Jacobi's theta functions.
These $Z^{(B1)}_{ab}$'s satisfy the modular properties 
in Eqs.(\ref{modular-invariance}).

By using these results, the total bosonic part of the partition function
$Z^{(B)}_{g_1,g_2}$ can be obtained by multiplying these
$Z^{(B1)}_{ab}$'s. For example, we take 
the $Z^{(B)}_{\alpha,\beta}$ concretely.
Since $\alpha$ and $\beta$ have the following actions
\begin{align*}
 \begin{array}{cccccccccc}
  \alpha:&( &-x^1, &-x^2, &-x^3, &-x^4,&x^5, &x^6, &x^7 &), \\
  \beta:&( &-x^1, &\dis\frac{1}{2}-x^2, &x^3, &x^4,&-x^5, &-x^6, &x^7 &), \\
 \end{array}
\end{align*}
the $x^1$-sector produces $Z^{(B1)}_{\2\2}$ and the $x^2$-sector
 produces $Z^{(B1)}_{\2\3}$ and so on. Consequently,
$Z^{(B)}_{\alpha,\beta}$ becomes a product of $Z^{(B1)}_{a,b}$'s for
each $x^i,\ (i=1,2,\dots,7)$
\begin{align*}
 Z^{(B)}_{\alpha,\beta}
=Z^{(B1)}_{\2\2}Z^{(B1)}_{\2\3}Z^{(B1)}_{\2\0}
Z^{(B1)}_{\2\0}Z^{(B1)}_{\0\2}Z^{(B1)}_{\0\2}Z^{(B1)}_{\0\0}.
\end{align*}

Next, we are going to write down $\chi^{E_8}_{g_1,g_2}$.
This part is essential for the spacetime gauge symmetry.
We use the description by free fermions, and the result can be
written by using five types of functions $\chi^{E_8}_{1,1}
,\chi^{E_8}_{1,\alpha},\chi^{E_8}_{\alpha,1},\chi^{E_8}_{\alpha,\alpha},
\chi^{E_8}_{\alpha,\gamma}$. 
We can write down their explicit formulae
by using theta functions
\begin{align}
& \chi^{E_8}_{1,1}=\chi^{E_8}_{\basi}=\frac1{2\eta^8}
\{\theta_3^8+\theta_4^8+\theta_2^{8}+(-i\theta_1)^8\},\nn\\
& \chi^{E_8}_{1,\alpha}=\frac1{2\eta^8}
\{\theta_3^6\theta_4^2+\theta_4^6\theta_3^2
-\theta_2^{6}(-i\theta_1)^2-(-i\theta_1)^6\theta_2^{2}\},\nn\\
& \chi^{E_8}_{\alpha,1}=\frac1{2\eta^8}
\{\theta_3^6\theta_2^2+\theta_2^6\theta_3^2
+\theta_4^{6}(-i\theta_1)^2+(-i\theta_1)^6\theta_4^{2}\},\nn\\
& \chi^{E_8}_{\alpha,\alpha}=\frac1{2\eta^8}
\{\theta_4^6\theta_2^2-\theta_2^6\theta_4^2
+\theta_3^{6}(-i\theta_1)^2-(-i\theta_1)^6\theta_3^{2}\},\nn\\
& \chi^{E_8}_{\alpha,\gamma}=\frac{i}{2\eta^8}
\{
\theta_3^5\theta_4\theta_2(-i\theta_1)+
\theta_2^5(-i\theta_1)\theta_3\theta_4+
\theta_4^5\theta_3(-i\theta_1)\theta_2+
(-i\theta_1)^5\theta_2\theta_4\theta_3
\}.
\label{chiE8-1}
\end{align}
The general $\chi^{E_8}_{g_1,g_2}$'s which are not defined in
Eqs.(\ref{chiE8-1}) are determined by the following rules
\begin{align*}
 \chi^{E_8}_{g_1,g_2}=
\begin{cases}
\chi^{E_8}_{1,\alpha} & (g_1=1,\ g_2\ne1)\\
\chi^{E_8}_{\alpha,1} & (g_1\ne1,\ g_2=1)\\
\chi^{E_8}_{\alpha,\alpha} & (g_1=g_2\ne1)\\
\chi^{E_8}_{\alpha,\gamma} & (g_1\ne g_2,\ g_1\ne 1,\ g_2\ne 1)\\
\end{cases},
\end{align*}
where the functions on the right-hand side are defined by Eqs.(\ref{chiE8-1}).

To check the modular invariance of the whole partition function, we need
the modular transformation properties of the above functions.
The modular properties of these functions are obtained by
using the modular properties of theta functions
in appendix \ref{appendix-theta}.
For S transformation, these $\chi^{E_8}$'s transform as
\begin{align}
& \chi^{E_8}_{1,1}(-1/\tau)=\chi^{E_8}_{1,1}(\tau),\quad
 \chi^{E_8}_{1,\alpha}(-1/\tau)=\chi^{E_8}_{\alpha,1}(\tau),\nn\\
& \chi^{E_8}_{\alpha,1}(-1/\tau)=\chi^{E_8}_{1,\alpha}(\tau),\quad
 \chi^{E_8}_{\alpha,\alpha}(-1/\tau)=-\chi^{E_8}_{\alpha,\alpha}(\tau),\quad
 \chi^{E_8}_{\alpha,\gamma}(-1/\tau)=
         \e{\frac14}\chi^{E_8}_{\gamma,\alpha}(\tau),
\label{modu1}
\end{align}
where we use $\e{x}:=\exp(2\pi i x)$.
On the other hand, for the T transformation, they transform as
\begin{align}
 & \chi^{E_8}_{1,1}(\tau+1)
  =\e{-\frac13}\chi^{E_8}_{1,1}(\tau),\quad
  \chi^{E_8}_{1,\alpha}(\tau+1)
  =\e{-\frac13}\chi^{E_8}_{1,\alpha}(\tau),\nn\\
 & \chi^{E_8}_{\alpha,1}(\tau+1)
  =\e{-\frac{1}{12}}\chi^{E_8}_{\alpha,\alpha}(\tau),\quad
\chi^{E_8}_{\alpha,\alpha}(\tau+1)
  =\e{-\frac{1}{12}}\chi^{E_8}_{\alpha,1}(\tau),\quad
\chi^{E_8}_{\alpha,\gamma}(\tau+1)
  =\e{-\frac{1}{12}}\chi^{E_8}_{\alpha,\alpha\gamma}(\tau).
\label{modu2}
\end{align}

Finally, we construct the left-moving fermionic part of the partition
function $Z^{(F)}_{g_1,g_2}$. This part is essential for the
spacetime supersymmetry.
We can construct this  $Z^{(F)}_{g_1,g_2}$ 
in the same way as $\chi^{E_8}_{g_1,g_2}$ case.
As constituent blocks, 
we need to write five types of partition functions
$Z^{(F)}_{1,1},Z^{(F)}_{1,\alpha},Z^{(F)}_{\alpha,1},
Z^{(F)}_{\alpha,\alpha},Z^{(F)}_{\alpha,\gamma}$. 
We can evaluate these functions concretely
and the results are
expressed as
\begin{align}
 &Z^{(F)}_{1,1}=
\frac{1}{2\eta^4}\{\theta_3^4-\theta_4^4-\theta_2^4+(-i\theta_1)^4\},\nn\\
 &Z^{(F)}_{1,\alpha}=
\frac{1}{2\eta^4}\{\theta_3^2\theta_4^2-\theta_4^2\theta_3^2
+\theta_2^2(-i\theta_1)^2-(-i\theta_1)^2\theta_2^2\},\nn\\
 &Z^{(F)}_{\alpha,1}=
\frac{1}{2\eta^4}\{\theta_3^2\theta_2^2-\theta_2^2\theta_3^2
-\theta_4^2(-i\theta_1)^2+(-i\theta_1)^2\theta_4^2\},\nn\\
 &Z^{(F)}_{\alpha,\alpha}=
\frac{1}{2\eta^4}\{-\theta_4^2\theta_2^2+\theta_2^2\theta_4^2
+\theta_3^2(-i\theta_1)^2-(-i\theta_1)^2\theta_3^2\},\nn\\
 &Z^{(F)}_{\alpha,\gamma}=
\frac{i}{\eta^4}\{
 \theta_3\theta_4\theta_2(-i\theta_1)
-\theta_2(-i\theta_1)\theta_3\theta_4
-\theta_4\theta_3(-i\theta_1)\theta_2
+(-i\theta_1)\theta_2\theta_4\theta_3
\}.
\label{ZF1}
\end{align}
Each of these functions actually vanishes because of
the spacetime supersymmetry.

The general $Z^{(F)}_{g_1,g_2}$'s which are not defined in
Eqs.(\ref{ZF1}) 
can be written as the same way as the $\chi^{E_8}_{g_1,g_2}$ case.
They are determined by the following rules
\begin{align*}
 Z^{(F)}_{g_1,g_2}=
\begin{cases}
Z^{(F)}_{1,\alpha} & (g_1=1,\ g_2\ne1)\\
Z^{(F)}_{\alpha,1} & (g_1\ne1,\ g_2=1)\\
Z^{(F)}_{\alpha,\alpha} & (g_1=g_2\ne1)\\
Z^{(F)}_{\alpha,\gamma} & (g_1\ne g_2,\ g_1\ne 1,\ g_2\ne 1)\\
\end{cases}.
\end{align*}

We also need the modular properties of these functions in
Eqs.(\ref{ZF1}) to check the modular invariance of the whole partition
function.  For the S transformation, they transform as
\begin{align}
& Z^{(F)}_{1,1}(-1/\tau)=Z^{(F)}_{1,1}(\tau),\quad
 Z^{(F)}_{1,\alpha}(-1/\tau)=Z^{(F)}_{\alpha,1}(\tau),\nn\\
& Z^{(F)}_{\alpha,1}(-1/\tau)=Z^{(F)}_{1,\alpha}(\tau),\quad
 Z^{(F)}_{\alpha,\alpha}(-1/\tau)=-Z^{(F)}_{\alpha,\alpha}(\tau),\quad
 Z^{(F)}_{\alpha,\gamma}(-1/\tau)=\e{\frac14}Z^{(F)}_{\gamma,\alpha}(\tau).
\label{modu3}
\end{align}
On the other hand, for the T transformation, they transform as
\begin{align}
 & Z^{(F)}_{1,1}(\tau+1)
  =\e{\frac13}Z^{(F)}_{1,1}(\tau),\quad
  Z^{(F)}_{1,\alpha}(\tau+1)
  =\e{\frac13}Z^{(F)}_{1,\alpha}(\tau),\nn\\
 & Z^{(F)}_{\alpha,1}(\tau+1)
  =\e{-\frac{5}{12}}Z^{(F)}_{\alpha,\alpha}(\tau),\quad
Z^{(F)}_{\alpha,\alpha}(\tau+1)
  =\e{-\frac{5}{12}}Z^{(F)}_{\alpha,1}(\tau),\quad
Z^{(F)}_{\alpha,\gamma}(\tau+1)
  =\e{-\frac{5}{12}}Z^{(F)}_{\alpha,\alpha\gamma}(\tau).\quad
\label{modu4}
\end{align}

Gathering these results, we can write down the $Z_{g_1,g_2}$ in 
Eq.(\ref{form-of-partition-function}).
Also we can check that the $Z_{g_1,g_2}$ 
actually satisfy the modular properties
(\ref{modular-invariance}) by using the modular properties 
(\ref{modu1}), (\ref{modu2}), (\ref{modu3}), (\ref{modu4})
 and we can conclude the partition function
is modular invariant.

\subsubsection{$Spin(7)$ holonomy manifold case}
Now, we turn to construct the modular invariant partition function of
the $Spin(7)$ example. It is almost parallel to the case of $G_2$.  In
$Spin(7)$ case, there are no transverse directions of the flat spacetime,
and $Z_{g_1,g_2}$ can be decomposed as 
\begin{align}
  &Z_{g_1,g_2}(\tau,\taub)=
Z^{(B)}_{g_1,g_2}(\tau,\taub)
\times Z^{(F)}_{g_1,g_2}(\tau)
\times \overline{(\chi^{E_8}_{g_1,g_2}(\tau)\chi^{E_8}_{\basi}(\tau))}.
\label{form-of-partition-function2}
\end{align}
The boson part $Z^{(B)}_{g_1,g_2}$ is constructed as in the $G_2$ case.
For example, $Z^{(B)}_{\alpha,\gamma}$ becomes a product of each $Z^{(B1)}$'s
\begin{align*}
 Z^{(B)}_{\alpha,\gamma}=
Z^{(B1)}_{\2\3}Z^{(B1)}_{\2\3}Z^{(B1)}_{\2\0}Z^{(B1)}_{\2\0}
Z^{(B1)}_{\0\3}Z^{(B1)}_{\0\3}Z^{(B1)}_{\0\0}Z^{(B1)}_{\0\0},
\end{align*}
where $Z^{(B1)}$'s are single boson partition functions
in Eqs.(\ref{boson-character}).

In order to write down the $Z^{(F)}_{g_1,g_2}$ and
$\chi^{E_8}_{g_1,g_2}$, let us note $\alpha\beta=:-1$.
We also write $(-1)\cdot g=-g$ for $\ g \in \Gamma$.
First, let us consider $\chi^{E_8}_{g_1,g_2}$. We need six new types of
functions which do not appear in Eqs.(\ref{chiE8-1}). These
functions are $\chi^{E_8}_{\alpha,-\alpha},
\chi^{E_8}_{\alpha,-1},\chi^{E_8}_{-1,\alpha},
\chi^{E_8}_{1,-1},\chi^{E_8}_{-1,1},\chi^{E_8}_{-1,-1}$.
The explicit forms of them are expressed as
\begin{align}
 &\chi^{E_8}_{\alpha,-\alpha}=\frac{1}{2\eta^8}\{
   \theta_3^4\theta_2^2\theta_4^2-\theta_2^4\theta_3^2(-i\theta_1)^2
   +\theta_4^4(-i\theta_1)^2\theta_3^2-(-i\theta_1)^4\theta_4^2\theta_2^2
   \},\nn\\
 &\chi^{E_8}_{\alpha,-1}=\frac{1}{2\eta^8}\{
   \theta_4^4\theta_2^2\theta_3^2+\theta_2^4\theta_4^2(-i\theta_1)^2
   +\theta_3^4(-i\theta_1)^2\theta_4^2+(-i\theta_1)^4\theta_3^2\theta_2^2
   \},\nn\\
 &\chi^{E_8}_{-1,\alpha}=\frac{1}{2\eta^8}\{
   \theta_2^4\theta_4^2\theta_3^2-\theta_4^4\theta_2^2(-i\theta_1)^2
   -\theta_3^4(-i\theta_1)^2\theta_2^2+(-i\theta_1)^4\theta_3^2\theta_4^2
   \},\nn\\
 &\chi^{E_8}_{1,-1}=\frac{1}{2\eta^8}\{
   \theta_3^4\theta_4^4+\theta_4^4\theta_3^4
   +\theta_2^4(-i\theta_1)^4+(-i\theta_1)^4\theta_2^4
   \},\nn\\
 &\chi^{E_8}_{-1,1}=\frac{1}{2\eta^8}\{
   \theta_3^4\theta_2^4+\theta_2^4\theta_3^4
   +\theta_4^4(-i\theta_1)^4+(-i\theta_1)^4\theta_4^4
   \},\nn\\
 &\chi^{E_8}_{-1,-1}=\frac{1}{2\eta^8}\{
   \theta_4^4\theta_2^4+\theta_2^4\theta_4^4
   +\theta_3^4(-i\theta_1)^4+(-i\theta_1)^4\theta_3^4
   \}.
\label{chiE8-2}
\end{align}

The general $\chi^{E_8}_{g_1,g_2}$'s which are not in
Eqs.(\ref{chiE8-1}), (\ref{chiE8-2}) are defined by using
the functions in Eqs.(\ref{chiE8-1}) and (\ref{chiE8-2}).
The general $\chi^{E_8}_{g_1,g_2}$ are defined as
\begin{align*}
 \chi^{E_8}_{g_1,g_2}=
\begin{cases}
\chi^{E_8}_{1,\alpha} & (g_1=1,\ g_2\ne1)\\
\chi^{E_8}_{-1,\alpha} & (g_1=-1,\ g_2\ne1)\\
\chi^{E_8}_{\alpha,1} & (g_1\ne1,\ g_2=1)\\
\chi^{E_8}_{\alpha,-1} & (g_1\ne1,\ g_2=-1)\\
\chi^{E_8}_{\alpha,\alpha} & (g_1=g_2\ne1)\\
\chi^{E_8}_{\alpha,-\alpha} & (g_1=-g_2\ne\pm1)\\
\chi^{E_8}_{\alpha,\gamma} & (\text{others}).\\
\end{cases}
\end{align*}

We also need the modular properties of six functions introduced in
Eq.(\ref{chiE8-2}) to check the modular invariance of
the whole partition function.
For the S transformation, they transform as
\begin{align}
 &\chi^{E_8}_{\alpha,-\alpha}(-1/\tau)=\chi^{E_8}_{-\alpha,\alpha}(\tau)
      ,\nn\\
 &\chi^{E_8}_{\alpha,-1}(-1/\tau)=\chi^{E_8}_{-1,\alpha}(\tau),\qquad
 \chi^{E_8}_{-1,\alpha}(-1/\tau)=\chi^{E_8}_{\alpha,-1}(\tau),\nn\\
 &\chi^{E_8}_{1,-1}(-1/\tau)=\chi^{E_8}_{-1,1}(\tau),\qquad
 \chi^{E_8}_{-1,1}(-1/\tau)=\chi^{E_8}_{1,-1}(\tau),\qquad
 \chi^{E_8}_{-1,-1}(-1/\tau)=\chi^{E_8}_{-1,-1}(\tau).\qquad
\label{modu5}
\end{align}
On the other hand, for the T transformation they behave as
\begin{align}
 &\chi^{E_8}_{\alpha,-\alpha}(\tau+1)
         =\e{-\frac{1}{12}}\chi^{E_8}_{\alpha,-1}(\tau),\qquad
  \chi^{E_8}_{\alpha,-1}(\tau+1)
         =\e{-\frac{1}{12}}\chi^{E_8}_{\alpha,-\alpha}(\tau),\nn\\
 &\chi^{E_8}_{-1,\alpha}(\tau+1)
         =\e{\frac{1}{6}}\chi^{E_8}_{-1,-\alpha}(\tau),\nn\\
 &\chi^{E_8}_{1,-1}(\tau+1)
         =\e{-\frac{1}{3}}\chi^{E_8}_{1,-1}(\tau),\qquad
  \chi^{E_8}_{-1,1}(\tau+1)
         =\e{\frac{1}{6}}\chi^{E_8}_{-1,-1}(\tau),\nn\\
 & \chi^{E_8}_{-1,-1}(\tau+1)
         =\e{\frac{1}{6}}\chi^{E_8}_{-1,1}(\tau).\qquad
\label{modu6}
\end{align}

As for the left-moving fermion part, we need six new types of
the functions.
We need explicit forms of $Z^{(F)}_{\alpha,-\alpha},
Z^{(F)}_{\alpha,-1},Z^{(F)}_{-1,\alpha},Z^{(F)}_{1,-1},
Z^{(F)}_{-1,1},Z^{(F)}_{-1,-1}$, and they are written as
\begin{align}
 &Z^{(F)}_{\alpha,-\alpha}=\frac{1}{2\eta^4}\{
      \theta_4^2\theta_2^2+(-i\theta_1)^2\theta_3^2
       -\theta_3^2(-i\theta_1)^2-\theta_2^2\theta_4^2
    \},\nn\\
 &Z^{(F)}_{\alpha,-1}=\frac{1}{2\eta^4}\{
      -\theta_3^2\theta_2^2-(-i\theta_1)^2\theta_4^2
       +\theta_4^2(-i\theta_1)^2+\theta_2^2\theta_3^2
    \},\nn\\
 &Z^{(F)}_{-1,\alpha}=\frac{1}{2\eta^4}\{
      -\theta_3^2\theta_4^2+(-i\theta_1)^2\theta_2^2
       -\theta_2^2(-i\theta_1)^2+\theta_4^2\theta_3^2
    \},\nn\\
 &Z^{(F)}_{1,-1}=Z^{(F)}_{-1,1}=\frac{1}{2\eta^4}\{
      \theta_4^4-\theta_3^4-(-i\theta_1)^4+\theta_2^4
    \},\nn\\
 &Z^{(F)}_{-1,-1}=\frac{1}{2\eta^4}\{
      \theta_3^4-\theta_4^4-\theta_2^4+(-i\theta_1)^4
    \}.\label{ZF2}
\end{align}

We introduce the general $Z^{(F)}_{g_1,g_2}$'s which are not in
 Eqs.(\ref{ZF1}) and (\ref{ZF2}). Each of these are the
same functions as in Eqs.(\ref{ZF1}),(\ref{ZF2}).
They can be defined as
\begin{align*}
 Z^{(F)}_{g_1,g_2}=
\begin{cases}
Z^{(F)}_{1,\alpha} & (g_1=1,\ g_2\ne1)\\
Z^{(F)}_{-1,\alpha} & (g_1=-1,\ g_2\ne1)\\
Z^{(F)}_{\alpha,1} & (g_1\ne1,\ g_2=1)\\
Z^{(F)}_{\alpha,-1} & (g_1\ne1,\ g_2=-1)\\
Z^{(F)}_{\alpha,\alpha} & (g_1=g_2\ne1)\\
Z^{(F)}_{\alpha,-\alpha} & (g_1=-g_2\ne\pm1)\\
Z^{(F)}_{\alpha,\gamma} & (\text{others}).\\
\end{cases}
\end{align*}

Here, we write down the modular properties of the functions in
Eqs.(\ref{ZF2}), which are needed to check the modular invariance of
the partition function.
For the S transformation, they transform as
\begin{align}
 &Z^{(F)}_{\alpha,-\alpha}(-1/\tau)=Z^{(F)}_{-\alpha,\alpha}(\tau),\nn\\
 &Z^{(F)}_{\alpha,-1}(-1/\tau)=Z^{(F)}_{-1,\alpha}(\tau),\qquad
 Z^{(F)}_{-1,\alpha}(-1/\tau)=Z^{(F)}_{\alpha,-1}(\tau),\nn\\
 &Z^{(F)}_{1,-1}(-1/\tau)=Z^{(F)}_{-1,1}(\tau),\qquad
 Z^{(F)}_{-1,1}(-1/\tau)=Z^{(F)}_{1,-1}(\tau),\qquad
 Z^{(F)}_{-1,-1}(-1/\tau)=Z^{(F)}_{-1,-1}(\tau).\qquad
\label{modu7}
\end{align}

For the T transformation, they behave as
\begin{align}
 &Z^{(F)}_{\alpha,-\alpha}(\tau+1)
         =\e{-\frac{5}{12}}Z^{(F)}_{\alpha,-1}(\tau),\qquad
  Z^{(F)}_{\alpha,-1}(\tau+1)
         =\e{-\frac{5}{12}}Z^{(F)}_{\alpha,-\alpha}(\tau),\nn\\
 &Z^{(F)}_{-1,\alpha}(\tau+1)
         =\e{-\frac{1}{6}}Z^{(F)}_{-1,-\alpha}(\tau),\nn\\
 &Z^{(F)}_{1,-1}(\tau+1)
         =\e{\frac{1}{3}}Z^{(F)}_{1,-1}(\tau),\qquad
  Z^{(F)}_{-1,1}(\tau+1)
         =\e{-\frac{1}{6}}Z^{(F)}_{-1,-1}(\tau),\nn\\
 & Z^{(F)}_{-1,-1}(\tau+1)
         =\e{-\frac{1}{6}}Z^{(F)}_{-1,1}(\tau).\qquad
\label{modu8}
\end{align}

The partition function constructed from these constituent blocks satisfies
the equations (\ref{modular-invariance}). It can be checked by using the
modular properties (\ref{modu1}), (\ref{modu2}), (\ref{modu3}),
(\ref{modu4}), (\ref{modu5}), (\ref{modu6}), (\ref{modu7}), 
(\ref{modu8}).

\subsection{massless sector}

In this subsection we will investigate massless spectra of the 
compactified models. 
The conformal dimension of a field in the whole theory is 
a sum of weights in each constituent CFT.
The total weight on the theory is labelled by a set
$(h^{tot},\bar{h}^{tot})$ and can be written down as
\ba
&&\mbox{left}\,(N=1)\,;\,{h}^{tot}={h}^M
+h^{\SO(d-2)}-\frac{12}{24}=0\,,\non
&&\mbox{right}\,(N=0)\,;\,\bar{h}^{tot}=\bar{h}^M
+\bar{h}^G+\bar{h}^{E_8}-\frac{24}{24}=0\,,\non
&&\rightarrow \bar{h}^M+\bar{h}^{G_0}+\bar{h}^{E_8}=1
\,,\,\,{h}^M+{h}^{\SO(d-2)}=\frac{1}{2}\,,\nom
\ea
where $(h^M,\bar{h}^M)$ expresses a set of weights in the extended CFT
for $M$ and $\bar{h}^{G_0}$, $\bar{h}^{E_8}$, $h^{\SO(d-2)}$ are 
respectively conformal dimensions associated with affine Lie
algebras $(\hat{G}_0)_1$, $(\hat{E}_8)_1$, $\hat{so}(d-2)_1$ $(d\geq 3)$. 
For the $d=3$ case we formally interpret the part ``$\hat{so}(d-2)_1$'' as 
a current generated by a free fermion. In the case of 
$d=2$ this part does not appear and we set $h^{\SO(d-2)}=0$.

As a first case we take gauge singlet states with conditions $(d\geq 3)$
\ba
\bar{h}^M=1\,,\,\,h^M+h^{\SO(d-2)}=\frac{1}{2}\,.\nom
\ea
The $h$'s are determined by representations of 
$\so(d-2)$ and can be classified in the following table \ref{t5}:\\
\begin{table}[htbp]
 \begin{enumerate}
 \item[] \underline{$d=even$ case\,\,}
 \ba
 \begin{array}{|c|cccc|}\hline
 \mbox{rep.} & \basi & \vect & \spi & \cospi \\\hline
 h^{\so(d-2)} & 0 & \frac{1}{2} & \frac{d-2}{16} &\frac{d-2}{16} \\\hline
 h^M & \frac{1}{2} & 0 & \frac{10-d}{16} & \frac{10-d}{16} \\\hline
 \mbox{sector} & NS & NS & R & R \\\hline
 \end{array}\nom
 \ea
 \item[] \underline{$d=odd$ case\,\,}
 \ba
 \begin{array}{|c|ccc|}\hline
 \mbox{rep.} & \basi & \vect & \spi  \\\hline
 h^{\so(d-2)} & 0 &  \frac{1}{2} & \frac{d-2}{16} \\\hline
 h^M & \frac{1}{2} & 0 & \frac{10-d}{16} \\\hline
 \mbox{sector} & NS & NS & R  \\\hline
 \end{array}\nom
\ea
\caption{Classifications of representations for $SO(d-2)$ algebra.}
\label{t5}
\end{enumerate}
\end{table}
In the table \ref{t5} we study models with spacetime transverse dimensions and 
the $NS$ and $R$ distinguish sectors of susy states
in the worldsheet theories.
For the $d=2$ case a condition $h^M=1/2$ should be satisfied.
By considering these conditions we can determine 
massless fields $(d>3)$ in this sector after GSO projections 
\ba
&&\psi^{\nu}_{-1/2}\tilde{\alpha}_{-1}^{\mu}\,;\,
\mbox{graviton},\,
\mbox{2nd rank antisymmetric field},\,\mbox{dilaton},\,\non
&&{S^{\alpha}}\tilde{\alpha}_{-1}^{\mu}\,;\,
\mbox{gravitino},\,\mbox{dilatino}\,.\nom
\ea
These represent an $N=1$ multiplet of supergravity.
For the $d=3$ case, the excitations of the gravity and gravitino disappear
after imposing on-shell conditions. 

In the $d=2$ case 
transverse dimension of the spacetime vanishes and 
local excitations of graviton and $B_{\mu\nu}$ do not exist.
However a pair of dilaton and dilatino appears as 
its field content.
For that case a set of weights is fixed to be 
$(h^M,\bar{h}^M)=(1,1/2)$ and could be classified 
by states of the CFT associated with the 
internal manifold $M^{(8)}$.

Secondly we consider the $\bar{h}^{E_8}=1$ part.
The corresponding states are easily understood to be 
gauge fields and their superpartners 
with gauge symmetry in the hidden sector $E_8$
\ba
&&\psi^{\mu}_{-1/2}\bar{J}^A_{-1}\,,\,\,
S^{\alpha}\bar{J}^A_{-1}\,,\,\,\non
&&\bar{J}^A\,;\,E_8\,\,\mbox{current}\,.\nom
\ea
These fields are singlets with respects to the 
$G_0$ group. 

In the case of $\bar{h}^{G_0}=1$ the corresponding states are 
gauge fields with spacetime visible gauge symmetry $G_0$. 
These transform as adjoint fields under this symmetry $G_0$ and
are identified with a set of an $N=1$ gauge multiplet 
\ba
&&\psi^{\mu}_{-1/2}\bar{J}^{\tilde{A}}_{-1}\,,\,\,
S^{\alpha}\bar{J}^{\tilde{A}}_{-1}\,,\,\,\non
&&\bar{J}^{\tilde{A}}\,;\,G_0\,\,\mbox{current}\,.\nom
\ea

Next we shall study the $E_8$ singlet matters with $h^{E_8}=0$. 
The right-moving part has an affine $G_0$ current and 
the states are classified by its representations. 
On the other hand the left-movers have $\SO(d-2)$ 
symmetry and its chiral states are labelled by 
representations of this group.
We will concentrate on the $d=2,3$ cases here.
The $d=3$ case is realized through compactification on the 
$G_2$ manifold with $D=7$.
The right- and left-chiral states are respectively characterized by 
representations of $G_0=\SO(9)$ and a free fermion $\psi $. 
They are summarized in the table \ref{t6}
\begin{table}[htbp]
 \ba
 &&\begin{array}{|c|ccccc|}\hline
 \SO(9) & \bar{h}^{\SO(9)} & 
 \bar{h}^{M} & (\bar{h}^{\tri},\bar{h}^r) & \mbox{tri-Ising} & \mbox{sector} \\\hline
 \basi\,\,(1) & 0 & 1 &  (3/5,2/5) & \epsilon' &  NS \\
 \vect\,\,(9) & 1/2 & 1/2 &  (1/10,2/5) & \epsilon &   NS \\
 \spi\,\,(16) & 9/16 & 7/16 &  (3/80,2/5) & \sigma &  R \\\hline
 \end{array}\,\nom
\ea
\caption{Right-moving part for $G_2$ case and its classification by $SO(9)$}
\label{t6}
\end{table}

Here the ``$\basi$'', ``$\vect$'', ``$\spi$'' express respectively trivial, vector, spinor 
representations of $\SO(9)$ and ``tri-Ising'' means 
scaling operators of the associated tricritical Ising model. 
Also the $\bar{h}^M$'s 
can be decomposed as sums of pairs of weights $(\bar{h}^{\tri},\bar{h}^r)$
of $(T^{\tri},T^r)$.
These states are collected into multiplets with a 
representation $26$ of $F_4$
\ba
 &&F_4\supset \SO(9)\times (\mbox{tricritical Ising})\,,\non
 &&26=1_{\basi}+9_{\vect}+16_{\spi}\,.\nom
\ea
The left-part is classified in terms of the 
transverse fermion $\psi$ in the spacetime.
The $h^M$'s are decomposed by the weights of the 
chiral fields of the tricritical Ising model as in table \ref{t7}

\begin{table}[htbp]
 \ba
 \begin{array}{|c|cccc|}\hline
 {h}^{\psi} & {h}^{M} & ( {h}^{\tri},{h}^r)&
 \mbox{tri-Ising} & \mbox{sector} \\\hline
  0 & 1/2 & (1/10,2/5) & \epsilon & NS \\
 1/2 & 0 & (0,0) & 1 &  NS \\
 1/16 & 7/16 & (3/80,2/5) & \sigma & R \\
 1/16 & 7/16 & (7/16,0) & \sigma' & R \\\hline
 \end{array}\,\nom
\ea
\caption{Left-moving part for $G_2$ case and its classification by $\psi$.}
\label{t7}
\end{table}
Now we are ready to write down spectra of the 
associated fields by gluing left- and right-parts
together. We put them in the table \ref{t8}:\\
\begin{table}[htbp]
 \ba
 \begin{array}{|c|c|c|c|c|c|c|c|}\hline
 \mbox{state} & \SO(9) & h^{\psi} & F_4 & \sharp\mbox{multiplet} &
 (h,\bar{h}) & ((-1)^F ,(-1)^{\bar{F}})
 \\\hline 
 (\frac{3}{5},\frac{2}{5})_L(\frac{1}{10},\frac{2}{5})_R & 1 & 0 &  &  &
 (1,\frac{1}{2}) & (+,-)\\
 (\frac{1}{10},\frac{2}{5})_L
 (\frac{1}{10},\frac{2}{5})_R & 9 & 0 & 26 & b_2+b_4 &
 (\frac{1}{2},\frac{1}{2}) & (-,-)\\
 (\frac{3}{80},\frac{2}{5})_L(\frac{1}{10},\frac{2}{5})_R & 16 & 0 &  &  &
 (\frac{7}{16},\frac{1}{2}) & (\pm ,-)\\\hline
 (\frac{3}{5},\frac{2}{5})_L(\frac{3}{80},\frac{2}{5})_R & 1 & 1/16 &  &  &
 (1,\frac{7}{16}) & (+,\pm )\\
 (\frac{1}{10},\frac{2}{5})_L
 (\frac{3}{80},\frac{2}{5})_R & 9 & 1/16 & 26 & b_2+b_4 &
 (\frac{1}{2},\frac{7}{16}) & (-,\pm )\\
 (\frac{3}{80},\frac{2}{5})_L(\frac{3}{80},\frac{2}{5})_R & 16 & 1/16 &  &  &
 (\frac{7}{16},\frac{7}{16}) & (\pm ,\pm )\\\hline
 \end{array}\nom
\ea
\caption{spectra ($d=3$  heterotic theory on $G_2$ manifold)}
\label{t8}
\end{table}
These states are $N=1$ $F_4$ fundamental multiplets and 
transform as a representation $26$ of gauge group $F_4$.
The number of these multiplets is 
evaluated by noticing the state 
$(\frac{3}{80},\frac{2}{5})_L(\frac{3}{80},\frac{2}{5})_R $. 
It is related with the string moduli space with $G_2$ manifold 
and its number is equal to the dimension of the ${\cal M}_{CFT}$, that
 is, $\dim {\cal M}_{CFT}=b_2+b_3=b_2+b_4$. 
In fact there are $b_2+b_3=55$ $F_4$ fundamental $26$-multiplets 
in our orbifold model. That illustrates the enhancement of the gauge
symmetry from $G_0=\SO(9)$ to $G=F_4$. 
Next we shall recall there are 
adjoint fields with a representation $36$ under $\SO(9)$. 
They are combined into an adjoint $52$-representation of $F_4$ 
together with $16$-matter fields of $\SO(9)$. 
Furthermore there are many gauge singlet states.
We do not touch on details of these singlets here.

When one compactifies string theory on the $Spin(7)$ manifold, 
the transverse dimension is $d-2=0$ 
and 
there are no transverse excitations. 
In our light-cone formula it seems meaningless to discuss 
matter contents for this case. 
But we will explain associated 
left- and right-parts formally for mathematical interests.
For simplicity we concentrate on the $\bar{h}^{E_8}=0$ sector. 
Then formal massless sectors are classified 
according to the representation of the gauge symmetry
$G_0=\SO(8)$ in the right-part. We show them in table \ref{t9}

\begin{table}[htbp]
 \ba
 &&\begin{array}{|c|cccc|}\hline
 \SO(8) & \bar{h}^{\SO(8)} & \bar{h}^{Spin(7)} &
 (\bar{h}^{\isi},\bar{h}^r) & \mbox{Ising}  \\\hline
 \basi\,\,(1) & 0 & 1 &  (1/2,1/2) & \epsilon  \\
 \vect\,\,(8_{\vect}) & 1/2 & 1/2 &  (0,1/2) & 1  \\
 \spi\,\,(8_{\spi}) & 1/2 & 1/2 &  (1/16,7/16) & \sigma  \\
 \cospi\,\,(8_{\cospi}) & 1/2 & 1/2 &  (1/16,7/16) & \sigma  \\\hline
 \end{array}\,\nom
\ea
\caption{Right-moving part for $Spin(7)$ case and its classification
 by $SO(8)$.}
\label{t9}
\end{table}
The $\basi$, $\vect$, $\spi$, $\cospi$ express respectively trivial, vector, spinor,
cospinor representations of $\SO(8)$ and ``Ising'' means 
scaling operators of the Ising model. 
The weights $\bar{h}^M$'s are decomposed into sums of pairs of 
$(\bar{h}^{\isi},\bar{h}^r)$. 
States here are collected into multiplets with 
representations $9$ and $16$ of an enhanced gauge symmetry $G=\SO(9)$
\ba
&&\SO(9)\supset \SO(8)\,,\non
&&9_{\vect}=1_{\basi}+8_{\vect}\,,\,\,16_{\spi}=8_{\spi}+8_{\cospi}\,.\nom
\ea
The $9_{\vect}$ and $16_{\spi}$ represent transformation 
properties of matter contents under the $\SO(9)$ 
and express respectively the vector and spinor representations of
$\SO(9)$. 

On the other side the left-part always has weights $h^{Spin(7)}=1/2$ and 
states are classified in terms of 
the chiral internal part in table \ref{t10}:

\begin{table}[htbp]
\ba
&&\begin{array}{|c|ccc|}\hline
{h} & 0 & 1/2 & 1 \\\hline
NS & \ket{0,0} & \ket{1/16,7/16} & \ket{1/2,1/2}\\
R &  & \ket{0,1/2}, \ket{1/2,0},\ket{1/16,7/16} & \\\hline
\end{array}\nom
\ea
\caption{Classification of chiral ground states for $Spin(7)$ case.}
\label{t10}
\end{table}
By gluing left- and right-parts together we 
can write down non-chiral states in the table \ref{t11}

\begin{table}[htbp]
 \ba
 \begin{array}{|c|c|c|c|c|c|c|}\hline
 \mbox{states} & \SO(8) & \SO(9) & \sharp\mbox{multiplet} &
 (h,\bar{h}) & ((-1)^F ,(-1)^{\bar{F}})
 \\\hline 
 \multicolumn{1}{|c|}{
 \begin{array}{c}
 (\frac{1}{2},\frac{1}{2})_{NS}
 (\frac{1}{16},\frac{7}{16})_{NS} \\
 (0,\frac{1}{2})_R(\frac{1}{16},\frac{7}{16})_{NS} 
 \end{array}}
 & 
 \multicolumn{1}{|c|}{%
 \begin{array}{c}
 1 \\
 8 
 \end{array}}
 & 
 \multicolumn{1}{|c|}{%
 \begin{array}{c}
 9 
 \end{array}}
 & 
 \multicolumn{1}{|c|}{%
 \begin{array}{c}
 b_3=b_5  
 \end{array}}
 &
 \multicolumn{1}{|c|}{%
 \begin{array}{c}
 (1,\frac{1}{2}) \\
 (\frac{1}{2},\frac{1}{2}) 
 \end{array}}
 & 
 \multicolumn{1}{|c|}{%
 \begin{array}{c}
 (+,-)\\
 (+,-)
 \end{array}}\\\hline
 (\frac{1}{16},\frac{7}{16})_{NS,R}(\frac{1}{16},\frac{7}{16})_{NS} & 8+8 & 16
 & 1+b_2+b_4^{-}  & (\frac{1}{2},\frac{1}{2}) & (- ,-)\\\hline
 \multicolumn{1}{|c|}{
 \begin{array}{c}
 (\frac{1}{2},\frac{1}{2})_{NS}(\frac{1}{16},\frac{7}{16})_{R} \\
 (0,\frac{1}{2})_R(\frac{1}{16},\frac{7}{16})_{R}
 \end{array}}
 & 
 \multicolumn{1}{|c|}{%
 \begin{array}{c}
 1 \\
 8 
 \end{array}}
 & 
 \multicolumn{1}{|c|}{%
 \begin{array}{c}
 9 
 \end{array}}
 & 
 \multicolumn{1}{|c|}{%
 \begin{array}{c}
 b_3=b_5  
 \end{array}}
 &
 \multicolumn{1}{|c|}{%
 \begin{array}{c}
 (1,\frac{1}{2}) \\
 (\frac{1}{2},\frac{1}{2}) 
 \end{array}}
 & 
 \multicolumn{1}{|c|}{%
 \begin{array}{c}
 (+,-)\\
 (+,-)
 \end{array}}\\\hline
 (\frac{1}{16},\frac{7}{16})_{NS,R}(\frac{1}{16},\frac{7}{16})_{R} & 8+8 & 16
 & 1+b_2+b_4^{-} & (\frac{1}{2},\frac{1}{2}) & (- ,-)\\\hline\hline
 \multicolumn{1}{|c|}{
 \begin{array}{c}
 (\frac{1}{2},\frac{1}{2})_{NS}(\frac{1}{2},\frac{1}{2})_{NS} \\
 (0,\frac{1}{2})_R(\frac{1}{2},\frac{1}{2})_{NS} 
 \end{array}}
 & 
 \multicolumn{1}{|c|}{%
 \begin{array}{c}
 1 \\
 8 
 \end{array}}
 & 
 \multicolumn{1}{|c|}{%
 \begin{array}{c}
 9 
 \end{array}}
 & 
 \multicolumn{1}{|c|}{%
 \begin{array}{c}
 b_6+b_4^{+}  
 \end{array}}
 &
 \multicolumn{1}{|c|}{%
 \begin{array}{c}
 (\frac{1}{2},1) \\
 (\frac{1}{2},{1}) 
 \end{array}}
 & 
 \multicolumn{1}{|c|}{%
 \begin{array}{c}
 (+,+)\\
 (+,+)
 \end{array}}\\\hline
 (\frac{1}{16},\frac{7}{16})_{NS,R}(\frac{1}{16},\frac{7}{16})_{NS} & 8+8 & 16
 & b_3=b_5  & (\frac{1}{2},{1}) & (- ,+)\\\hline
 \multicolumn{1}{|c|}{
 \begin{array}{c}
 (\frac{1}{2},\frac{1}{2})_{NS}(0,\frac{1}{2})_{R} \\
 (0,\frac{1}{2})_R(0,\frac{1}{2})_{R} 
 \end{array}}
 & 
 \multicolumn{1}{|c|}{%
 \begin{array}{c}
 1 \\
 8 
 \end{array}}
 & 
 \multicolumn{1}{|c|}{%
 \begin{array}{c}
 9 
 \end{array}}
 & 
 \multicolumn{1}{|c|}{%
 \begin{array}{c}
 b_6+b_4^{+}  
 \end{array}}
 &
 \multicolumn{1}{|c|}{%
 \begin{array}{c}
 (\frac{1}{2},\frac{1}{2}) \\
 (\frac{1}{2},\frac{1}{2}) 
 \end{array}}
 & 
 \multicolumn{1}{|c|}{%
 \begin{array}{c}
 (+,+)\\
 (+,+)
 \end{array}}\\\hline
 (\frac{1}{16},\frac{7}{16})_{NS,R}(0,\frac{1}{2})_{R} & 8+8 & 16
 & b_3=b_5  & (\frac{1}{2},\frac{1}{2}) & (- ,+)\\\hline
 \end{array}\nom
\ea
\caption{ spectra ($d=2$  heterotic theory on $Spin(7)$ manifold)}
\label{t11}
\end{table}
These states are $N=1$ $\SO(9)$ gauge multiplets and 
transform as representations $9$ (vector) and $16$ (spinor). 
The number of these multiplet is evaluated by comparing the states 
in Eq.(\ref{sp7}). 
It is related with the Betti numbers of $Spin(7)$ manifold and 
they are evaluated in our orbifold model as
\ba
&&b_0=b_8=1\,,\,\,b_1=b_7=0\,,\non
&&b_2=b_6=12\,,\,b_3=b_5=16\,,\,b_4=150\,,\,\non
&&b_4^{+}=107\,,\,b_4^{-}=43\,.\nom
\ea
By using these data, we can calculate multiplicities of the $\SO(9)$ 
matters
\ba
&&\{\mbox{multiplicity of $9$}\}
=b_3=b_5=16\,,\non
&&\{\mbox{multiplicity of $9$}\}=b_6+b_4^{+}=119\,,\non
&&\{\mbox{multiplicity of $16$}\}=b_3=b_5=16\,,\non
&&\{\mbox{multiplicity of $16$}\}=1+b_2+b_4^{-}=56\,.\nom
\ea
In particular the multiplet of representation $16$ 
has multiplicity $1+b_2+b_4^{-}=56$. 
It coincides with the dimension of the 
string moduli space ${\cal M}_{CFT}$.

\section{Conclusions and Discussions}

In this paper, we investigated heterotic strings on the 
exceptional holonomy manifolds by making use of the 
CFT techniques and found a cascade of special holonomy manifolds 
with different dimensions. 
In order to analyze these phenomena, we used standard CFT
techniques of 
branching rules for characters.

We study partition functions of $E_8\times E_8$ 
heterotic strings compactified on these manifolds and find that gauge
symmetry enhancements are correlated with reductions 
of holonomies of the internal manifolds. 

Gauge symmetry parts are exceptional groups $E_6$, $F_4$ respectively 
for $CY_3$ and $G_2$ theories and the $Spin(7)$ theory has an $\SO(9)$ 
gauge symmetry in $2$ dim spacetime.

The criticality condition on the left-moving side
of this superstring is equivalent 
to a relation 
$d+D=10$ for dimensions $d$, $D$ of spacetime and 
internal parts.
In addition there are conditions for
central charges $c_G$, $c_{hol}$, $c_{spec}$
 in the gauge sector 
on the right-moving side 
$c_G+c_{hol}=8$, $D=2(c_{hol}+c_{spec})$. 
The $c_G$, $c_{hol}$
correspond to (enhanced) gauge group
$G$, 
holonomy group $G_{hol}$ and the central charge $c_{spec}$ is associated
with a CFT of a spectral flow operator. 
They also give us information on division of $E_8$ into 
the holonomy group $G_{hol}$ and (enhanced) gauge group $G$. 

The essential part of our mechanism  
originates in two equations 
about characters 
$\chi^{E_8}=\chi^{\SO(8)}\times \chi^{\SO(8)}$, 
$\chi^{\SO(8)/\SU(3)}=\chi^{\isi}\times \chi^{\tri}\times
\chi^{\pot}=
\chi^{\SO(2)}\times \chi^{U(1)}$. 
By multiplying each character of $\chi^{\isi}\times \chi^{\tri}\times
\chi^{\pot}$
one after another to the $\SO(8)$ part, 
we can obtain characters of visible enhanced gauge symmetries
$\SO(9)$, $F_4$, $E_6$.
At the same time holonomy parts are reduced to 
$Spin(7)$, $G_2$, $\SU(3)$ and 
 associated manifolds could be 
changed. 
The first is the $Spin(7)$ holonomy case. 
The holonomy $\SO(8)$ part is decomposed into 
$Spin(7)$ in terms of 
Ising model because of an equation 
$\chi^{\SO(8)}=\chi^{Spin(7)}\times \chi^{\isi}$. 
The second is a reduction from the 
$Spin(7)$ to $G_2$ 
holonomy 
by throwing away degrees of freedom of 
the tricritical Ising model. 
It can be explained by an equation 
$\chi^{Spin(7)}=\chi^{G_2}\times \chi^{\tri}$. 
This statistical model with $c=7/10 $
acts on the gauge part of $\SO(9)$ 
and lifting it to the new symmetry $F_4$ 
through an equation 
$\chi^{\SO(9)}\times \chi^{\tri}=\chi^{F_4}$. 
The last comes from a 
relation 
$\chi^{G_2}=\chi^{\SU(3)}\times \chi^{\pot}$
including 3-state Potts model. 
It explains a reduction of holonomy from $G_2$ to 
$\SU(3)$, that is, 
a relation of $G_2$ manifolds and Calabi-Yau $3$-folds. 
It also changes gauge symmetries 
in spacetimes from $F_4$ to $E_6$ 
because we have a relation 
$\chi^{F_4}\times \chi^{\pot}=\chi^{E_6}$. 
By noticing the relation $\chi^{\SO(8)/\SU(3)}=\chi^{\isi}\times
\chi^{\tri}\times \chi^{\pot}$, 
we can understand the associated CFT 
has an affine $U(1)$ symmetry 
needed to enhance 
the worldsheet $N=1$ CFT algebra to 
$N=2$ conformal algebra of $CY_3$. 
It can be explained 
by an identity
$\chi^{\isi}\times \chi^{\tri}\times
\chi^{\pot}=
\chi^{\SO(2)}\times \chi^{U(1)}$. 
At the level of balance of central charges, 
this equation means relations 
$c=\frac{1}{2}+\frac{7}{10}+\frac{4}{5}=2=1+1$.

This $U(1)$ serves as an R-symmetry of the 
$N=2$ theory and 
is used to construct $U(1)$ current of the 
$N=2$ algebra.
Also a spectral flow operator of the $N=2$ CFT 
has conformal dimension $3/8$ 
and is constructed 
by combining 
scaling operators of three statistical models. 
It is realized as a combination of states 
with $(h^{\isi},h^{\tri},h^{\pot})=(1/16,7/16,0),(1/16,3/80,2/5)$. 
This operator belongs to a sector 
with a $U(1)$ charge $Q=3/2$. 
It implies that the state is related with 
a $3$-form of the $CY_3$.
Also the total weight of these states turns out to be $1/2$ and 
we can obtain a spin operator $\Sigma$ with $h=3/8(=1/2-1/8)$ of the 
$\SU(3)$ holonomy model 
by subtracting contributions of a spin operator of $\SO(2)$ 
with weight $1/8$.
This operator $\Sigma$ 
is nothing but a holomorphic $3$-form of the $CY_3$ and 
confirms the validity of our discussions. 

By using this operator $\Sigma$ and 
combining the 4dim spacetime spin operator $S_{\alpha}$ together with
contribution of ghost part, 
we can construct 
a spacetime supercharge $Q_{\alpha}=\int
e^{-\frac{1}{2}\phi}S_{\alpha}\Sigma$. It 
guarantees spacetime ${\cal N}=1$ supersymmetry.

It is amazing that we can realize 
$N=2$ CFT associated with $CY_3$ starting from $\SO(8)$ theory 
by using three statistical models in $2$ dimension. 
We will make several comments about these models here:The Ising model 
appears as the first entry (that is, with the lowest central charge) of minimal unitary models 
with $N=0$. The tricritical Ising is a second model (with the lowest $c$
but one) in the $N=0$ minimal series. 
But it is also a model in the $N=1$ unitary minimal model 
with the lowest central charge. 
Furthermore the 3-state Potts model is the third model 
in the $N=0$ minimal series but it is the first model 
in a series of the $W_3$ algebra. 
It is a challenging task to analyze more precisely these structures.
Particularly it is known that extended $N=2$ algebras of $CY_3$'s 
have $W$-like symmetries, so-called 
$c=9$ algebras. These structures with higher spin currents 
might be related with the 
$W_3$ algebra of the 3-state Potts model. 

In our heterotic theory, the left-part has worldsheet $N=1$ 
supersymmetry. 
This left-sector is composed of 
the internal manifold and transverse Lorentz group $\SO(d-2)$.
Owing to the supersymmetry the left-part of the toroidal partition
function vanishes 
by using 
identities
about theta functions. 
We propose
an identity that guarantees this symmetry
in the context of CFTs for these minimal models (Ising, tricritical
Ising, 3-state Potts). 
This left-part has the Lorentz group $\SO(d-2)$ and 
it contains information about 
spacetime dimension $d$. 
By changing holonomy groups 
there appear 
identities for
characters associated with internal manifolds. 
They are some kinds of theta identities and 
could explain the dimension $d$
through some balance with the $\SO(d-2)$ part.

We would like to emphasize that our 
results are obtained under the completely general backgrounds. 
Especially it is remarkable that the forms of characters of 
statistical models are perfectly fitted to the 
holonomy parts of the manifolds in the gauge sector 
of the partition functions. 
Moreover 
identities in the worldsheet susy part are 
related with 
transverse Lorentz groups $\SO(d-2)$
combined with characters of CFTs for these 
special manifolds.

In section 5,
we take concrete examples 
realized as orbifolds discussed by Joyce.
We construct toroidal partition functions 
of heterotic strings 
compactified on these exceptional manifolds. 
We analyzed properties under modular transformations 
and studied consistencies of the 
strings on the orbifolds. 
Also we elaborate the 
spectra of massless sector of these models. 
For the $G_2$ case the matter parts are classified by 
representations of the gauge group $F_4$ and 
they are collected into $3$ dim $N=1$ multiplets of 
an $F_4$ gauge (supergravity) theory. 
The fundamental multiplets with $26$-representation 
of $F_4$ are related with the (string)
moduli space ${\cal M}_{CFT}$ of the internal $G_2$ manifold
and its multiplicity is evaluated by a combination of topological numbers
$b_2+b_3=55$.

In the case of the $Spin(7)$ manifold, 
the matter parts transform as $9$- and $16$-representations under the
enhanced gauge symmetry $\SO(9)$. 
The associated fields of massless sectors 
are collected into $2$ dim $N=1$ multiplets 
of an $\SO(9)$ gauge (supergravity) theory. 
The multiplets with 
spinor $16$-representation 
of $\SO(9)$ 
correspond to cohomology elements of
the (string) moduli space ${\cal M}_{CFT}$ 
of the $Spin(7)$ manifold. 
Its multiplicity is calculated by using topological numbers 
as $1+b_2+b_4^{-}=56$.

\subsection*{Acknowledgement}
The authors would like to thank Kentaroh Yoshida
for useful discussions and comments.  S.Y. would
also like to thank the organizers of the Summer
Institute 2001 at Yamanashi, Japan, 6-20 August, 2001, where a part of
this work is done.

The work of S.Y. is supported in part by the JSPS Research Fellowships for
Young Scientists.


\appendix
\section{Theta functions}\label{appendix-theta}
We will review some properties of theta functions.
The theta function is defined as
\ba
&&\vth(\nu |\tau)=\sum_{n\in\bz}e^{\pi in^2\tau +2\pi in\nu}\,.\nom
\ea
The Jacobi's triple product identity is expressed in the following formula
\ba
&&\vth(\nu |\tau)=\prod_{n=1}^{\infty}(1-q^n)
(1+zq^{n-\frac{1}{2}})(1+z^{-1}q^{n-\frac{1}{2}})
\,,\qquad
q=e^{2\pi i\tau}\,,\,z=e^{2\pi i\nu}\,.\nom
\ea
This function has periodicity $1$ and its
modular properties are summarized as
\ba
&&\vth(\nu+1 |\tau)=\vth (\nu|\tau)\,,\qquad
\vth (\nu +\tau |\tau)=e^{-\pi i\tau -2\pi i\nu}
\vth (\nu |\tau)\,,\non
&&\vth (\nu |\tau +1)=\vth (\nu +\frac{1}{2}|\tau)\,,\qquad
\vth (\nu/ \tau|-1/ \tau)
=(-i\tau)^{1/2}e^{\pi i\frac{\nu^2}{\tau}}\vth (\nu |\tau)\,.\nom
\ea
 Generalized theta functions are defined as
\ba
&&\vthe{a}{b}(\nu |\tau )=
\sum_{n\in \bz}\exp 
\left[ \pi i (n+a)^2 \tau +2\pi i (n+a)(\nu +b)\right]\non
&&\qquad \qquad =
\exp \left[
      \pi i a^2 \tau +2\pi i a (\nu +b)
     \right]\vth (\nu +a\tau +b ,\tau)\,.\nom
\ea
Ordinary Jacobi's theta functions are defined by using the
generalized theta function
\ba
&&\vth_3 (\nu|\tau)
=\vthe{0}{0}(\nu |\tau)=\sum_{n\in\bz}
q^{\frac{1}{2}n^2}z^n\,,\qquad
\vth_1 (\nu|\tau)
=-\vthe{1/2}{1/2}(\nu |\tau)=i \sum_{n\in\bz}
(-1)^nq^{\frac{1}{2}(n-\frac{1}{2})^2}z^{n-\frac{1}{2}}\,,\non
&&\vth_2 (\nu|\tau)
=\vthe{1/2}{0}(\nu |\tau)=\sum_{n\in\bz}
q^{\frac{1}{2}(n-\frac{1}{2})^2}z^{n-\frac{1}{2}}\,,\qquad
\vth_4 (\nu|\tau)
=\vthe{0}{1/2}(\nu |\tau)=\sum_{n\in\bz}
(-1)^nq^{\frac{1}{2}n^2}z^n\,.\nom
\ea
These theta functions are also expressed as infinite products
\ba
&&\vth_{3}(\nu |\tau)=\prod_{n=1}^{\infty}
(1-q^n)(1+zq^{n-\frac{1}{2}})(1+z^{-1}q^{n-\frac{1}{2}})\,,\non
&&\vth_{4}(\nu |\tau)=\prod_{n=1}^{\infty}
(1-q^n)(1-zq^{n-\frac{1}{2}})(1-z^{-1}q^{n-\frac{1}{2}})\,,\non
&&\vth_{2}(\nu |\tau)=2e^{\frac{\pi i}{4}\tau}\cos \pi \nu
\prod_{n=1}^{\infty}
(1-q^n)(1+zq^{n})(1+z^{-1}q^{n})\,,\non
&&\vth_{1}(\nu |\tau)=-2e^{\frac{\pi i}{4}\tau}\sin \pi \nu
\prod_{n=1}^{\infty}
(1-q^n)(1-zq^{n})(1-z^{-1}q^{n})\,.\nom
\ea
The Dedekind eta function is frequently used
\ba
&&\eta (\tau)=q^{\frac{1}{24}}
\prod_{n=1}^{\infty}(1-q^n)\,.\nom
\ea
The modular properties of these functions are important and shown in
the 
following equations
\ba
&&\vth_{3}(\nu |\tau +1)=\vth_4 (\nu |\tau)\,,\qquad
\vth_{4}(\nu |\tau +1)=\vth_3 (\nu |\tau)\,,\non
&&\vth_{2}(\nu |\tau +1)=e^{\frac{\pi i}{4}}\vth_2 (\nu |\tau)\,,\qquad
\vth_{1}(\nu |\tau +1)=e^{\frac{\pi i}{4}}\vth_1 (\nu |\tau)\,,\qquad
\eta (\tau +1)=e^{\frac{\pi i}{12}}\eta (\tau)\,,\non
&&\vth_3(\frac{\nu}{\tau} |\frac{-1}{\tau})=
(-i\tau)^{\frac12}e^{\frac{\pi i}{\tau}\nu^2}\vth_3(\nu |\tau)\,,\qquad
\vth_4(\frac{\nu}{\tau} |\frac{-1}{\tau})=
(-i\tau)^{\frac12}e^{\frac{\pi i}{\tau}\nu^2}\vth_2(\nu |\tau)\,,\non
&&\vth_2(\frac{\nu}{\tau} |\frac{-1}{\tau})=
(-i\tau)^{\frac12}e^{\frac{\pi i}{\tau}\nu^2}\vth_4(\nu |\tau)\,,\qquad
\vth_1(\frac{\nu}{\tau} |\frac{-1}{\tau})=
(-i\tau)^{\frac12}e^{\frac{\pi i}{\tau}\nu^2}\vth_1(\nu |\tau)\,,\non
&&\eta (-\frac{1}{\tau})=(-i\tau)^{\frac12}\eta (\tau)\,.\nom
\ea
We also use the classical SU(2) theta function defined as
\begin{align*}
 \Theta_{m,k}(\nu|\tau)=\sum_{n\in\Zb}q^{k\left(n+\frac{m}{2k}\right)^2}
                        z^{k\left(n+\frac{m}{2k}\right)}.
\end{align*}
We sometimes abbreviate arguments of theta functions: For example,
$\theta_3=\theta_3(\tau)$ means $\theta_3(\nu=0|\tau)$.

\section{CFT and characters}
\subsection{Minimal models}
The unitary minimal models are labeled by an integer $m$
($m=3,4,5,\dots$). 
Its central
charge is given by a formula
\begin{align*}
 c=1-\frac{6}{m(m+1)}.
\end{align*}
The Verma modules of each minimal model is classified by integers $r,s$ in the
regions
\begin{align*}
 r=1,2, \dots,m-1,\qquad s=1,2, \dots,m,\qquad \mbox{with}\,\,ms<(m+1)r.
\end{align*}
The conformal dimension of
the primary field is specified by the set $(r,s)$ and is evaluated as
\begin{align*}
 h_{r,s}=\frac{\{(m+1)r-ms\}^2-1}{4m(m+1)}.
\end{align*}
The characters of these minimal models can be expressed for the 
primary field labelled by $(r,s)$
\begin{align*}
 \chi^{(m)}_{r,s}=\frac{1}{\eta(\tau)}
\{\Theta_{(m+1)r-ms,m(m+1)}(\tau)-\Theta_{(m+1)r+ms,m(m+1)}(\tau)\}.
\end{align*}
We use $m=3,4,5$ minimal models in this paper. The details of properties
of these models are listed in the following table:
\begin{itemize}
 \item \underline{Ising model\,\,} \qquad ($c=\frac12$)
 \begin{align*}
  h_{1,1}=0,\ h_{2,1}=\frac12,\ h_{1,2}=\frac1{16}.
 \end{align*}
 We write the Virasoro characters for this model as $\chi^{\isi}_{h_{r,s}}$.
 \item \underline{Tricritical Ising model\,\,} \qquad ($c=\frac{7}{10}$)
 \begin{align*}
  h_{1,1}=0,\ h_{2,1}=\frac{7}{16},\ h_{1,2}=\frac 1{10},\ 
h_{1,3}=\frac 35, \ h_{2,2}=\frac 3{80},\ h_{3,1}=\frac 32.
 \end{align*}
 We write the Virasoro characters of this model as $\chi^{\tri}_{h_{r,s}}$.
 \item \underline{3-state Potts model\,\,} \qquad ($c=\frac{4}{5}$)
 \begin{align*}
  h_{1,1}=0,\ h_{2,1}=\frac{2}{5},\ h_{3,1}=\frac 75,\ 
h_{1,3}=\frac 23, \ h_{4,1}=3,\ h_{2,3}=\frac 1{15}.
 \end{align*}
The notation $\chi^{\pot}_{h_{r,s}}$ is used for Virasoro characters 
for this Potts model. But we 
mainly use $W_3$ characters constructed from those of the 
Potts model
\begin{align*}
& C^{\pot}_{0}=\chi^{\pot}_{0}+\chi^{\pot}_{3},\quad
& C^{\pot}_{2/5}=\chi^{\pot}_{2/5}+\chi^{\pot}_{7/5}, \\
& C^{\pot}_{2/3}=\chi^{\pot}_{2/3},\quad
& C^{\pot}_{1/15}=\chi^{\pot}_{1/15}.\quad
\end{align*}
The standard modular invariant partition function of the 3-state Potts model
can be described by using these $W_3$ characters $C^{\pot}$'s 
\begin{align*}
 Z=|C^{\pot}_{0}|^2+|C^{\pot}_{2/5}|^2+2|C^{\pot}_{2/3}|^2
+2|C^{\pot}_{1/15}|^2.
\end{align*}
\end{itemize}
\subsection{WZW models}
\label{appendix-WZW}
The central charges and the conformal dimensions 
are summarized 
for representations
of level 1 affine Lie algebras 
in table \ref{wzw}
\begin{table}[htbp]
 \begin{tabular}{|c||c||c|c|c|}\hline
 group & center & \basi &\fund & $\cfund$ \\ \hline\hline
 \su(2) & $1$ & $0$ & $1/4$  & $-$ \\ \hline
 \su(3) & $2$ & $0$ & $1/3$  & $1/3$ \\ \hline
 $G_2$ & $14/5$ & $0$ & $2/5$  & $-$ \\ \hline
 $F_4$ & $26/5$ & $0$ & $3/5$  & $-$ \\ \hline
 $E_6$ & $6$ & $0$ & $2/3$  & $2/3$ \\ \hline
 $E_7$ & $7$ & $0$ & $3/4$  & $-$ \\ \hline
 $E_8$ & $8$ & $0$ & $-$  & $-$ \\ \hline
 \end{tabular}
 \begin{tabular}{|c||c||c|c|c|c|}\hline
 group & center & \basi & \vect &\spi & $\cospi$ \\ \hline\hline
  \so($2r$) & $r$ &$0$ &$1/2$ &$r/8$ &$r/8$ \\ \hline
  \so($2r+1$) & $r+1/2$ &$0$ &$1/2$ &$(2r+1)/16$ & $-$ \\ \hline
 \end{tabular}
 \caption{Properties of level 1 affine Lie algebras. The central charge
 and conformal dimension of each representation is shown here.  The
 symbol ``$-$'' means there are no such representations. } 
\label{wzw}
\end{table}

Explicit forms of characters used in this paper are written down as follows
\ba
&&\chi^{\so(2r)}_{\basi}
         =\frac12\left(\left(\frac{\theta_3}{\eta}\right)^{r}
              +\left(\frac{\theta_4}{\eta}\right)^{r}\right),\qquad
\chi^{\so(2r)}_{\vect}
         =\frac12\left(\left(\frac{\theta_3}{\eta}\right)^{r}
              -\left(\frac{\theta_4}{\eta}\right)^{r}\right),\nn\\
&&\chi^{\so(2r)}_{\spi}=\chi^{\so(2r)}_{\cospi}
      =\frac1{2}\left(\frac{\theta_2}{\eta}\right)^{r},
\qquad \chi^{\so(2r+1)}_{\spi}
  =\frac1{\sqrt2}\left(\frac{\theta_2}{\eta}\right)^{\frac{2r+1}{2}},\nn\\
&&\chi^{\so(2r+1)}_{\basi}
       =\frac12\left(\left(\frac{\theta_3}{\eta}\right)^{\frac{2r+1}{2}}
       +\left(\frac{\theta_4}{\eta}\right)^{\frac{2r+1}{2}}\right),
\qquad\chi^{\so(2r+1)}_{\vect}
       =\frac12\left(\left(\frac{\theta_3}{\eta}\right)^{\frac{2r+1}{2}}
       -\left(\frac{\theta_4}{\eta}\right)^{\frac{2r+1}{2}}\right),\nn\\
&&  \chi^{\su(2)}_{\basi}=\frac{\Theta_{0,1}}{\eta},\qquad
   \chi^{\su(2)}_{\fund}=\frac{\Theta_{1,1}}{\eta},\nn\\
&&    \chi^{\su(3)}_{\basi}=
      \frac1{\eta^2}\left(
         \Theta_{0,3}\Theta_{0,1}+\Theta_{3,3}\Theta_{1,1}
      \right),\qquad
   \chi^{\su(3)}_{\fund}=\chi^{\su(3)}_{\cfund}=
      \frac1{\eta^2}\left(
         \Theta_{2,3}\Theta_{0,1}+\Theta_{1,3}\Theta_{1,1}
      \right),\nn\\
 &&\chi^{E6}_{\basi}=
 \frac{1}{2\eta^6(\tau)}\left\{
 \theta_3(3\tau)\cdot \theta_3(\tau)^5+
 \theta_4(3\tau)\cdot \theta_4(\tau)^5+
 \theta_2(3\tau)\cdot \theta_2(\tau)^5
 \right\}\,,\non
 &&\chi^{E6}_{\fund}=
 \frac{1}{2\eta^6(\tau)}\Bigl\{
 \vthe{1/6}{0}(3\tau)\cdot \theta_2(\tau)^5
 + 
 \vthe{2/3}{0}(3\tau)\cdot \theta_3(\tau)^5+
 e^{-2\pi i/3}
 \vthe{2/3}{1/2}(3\tau)\cdot \theta_4(\tau)^5
 \Bigr\}\,,\non
 &&\chi^{E6}_{\cfund}=
 \frac{1}{2\eta^6(\tau)}\Bigl\{
 \vthe{5/6}{0}( 3\tau)\cdot \theta_2( \tau)^5
 +
 \vthe{1/3}{0}( 3\tau)\cdot \theta_3( \tau)^5-
 e^{-\pi i/3}\vthe{1/3}{1/2}( 3\tau)\cdot \theta_4( \tau)^5
 \Bigr\}\,,\non
 &&\chi^{E7}_{\basi}=
 \frac{1}{2\eta^7(\tau)}\left\{
 \theta_2( 2\tau)\cdot \theta_2( \tau)^6+
 \theta_3( 2\tau)\cdot\left( \theta_3( \tau)^6+
 \theta_4( \tau)^6\right)
 \right\}\,,\non
 &&\chi^{E7}_{\fund}=
 \frac{1}{2\eta^7(\tau)}\left\{
 \theta_3( 2\tau)\cdot \theta_2( \tau)^6+
 \theta_2( 2\tau)\cdot\left( \theta_3( \tau)^6-
 \theta_4( \tau)^6\right)
 \right\}\,,\non
 &&\chi^{E8}_{\basi}=
 \frac{1}{2\eta^8(\tau)}\left\{
 \theta_2( \tau)^8
 +\theta_3( \tau)^8
 +\theta_4( \tau)^8
 \right\}\,.\nom
\ea
The algebra $\uone_k$, ($ k\in \Zb,\ k>0$) also appears. 
Each module of the $\uone_k$ is labeled by an integer
$m\in \Zb_{2k}$, and a character of a module $m$ is $\Theta_{m,k}/\eta$.
The partition function of this CFT is written as
\begin{align*}
 Z=\sum_{m\in\Zb_{2k}}|\Theta_{m,k}/\eta|^2.
\end{align*}
This theory describes a single free boson of radius $\sqrt{2k}$. 
We make a remark here:
the $\uone_1$ is the level 1 $\su(2)$ algebra, and $\uone_2$
represents a level 1 affine $\so(2)$ algebra in our notation.

\providecommand{\href}[2]{#2}\begingroup\raggedright\endgroup

\end{document}